\documentclass[copyright,creativecommons]{eptcs}

\providecommand{\event}{EXPRESS/SOS 2024}

\usepackage[utf8]{inputenc}
\usepackage{bold-extra}
\usepackage{amsmath,
	    amsfonts,
	    amstext,
	    amssymb,
	    mathrsfs,
	    mathpartir,
	    yhmath,
	    stmaryrd,
	    latexsym,
	    enumitem,
	    proof,
	    graphicx,
	    color,
	    hyperref,
	    comment,
	    semantic,
	    float,
	    hhline,
	    tikz}

\newcommand{\ms}[1]
	{\null\ifmmode\mathord{\mathcode`-="702D\it #1\mathcode`\-="2200}
	\else$\mathord{\mathcode`-="702D\it #1\mathcode`\-="2200}$\fi}

\newcommand{\cws}[2]
	{\\ \centerline{$#2$} \\[-#1pt]}

\newcommand{\fullbox}
	{{\mbox{}\nolinebreak\hfill{$\rule{1.5mm}{1.5mm}$}}}

\newcommand{\bibtrick}[1]
	{}

\newcommand{\lap}
	{\mbox{$<$}}
\newcommand{\rap}
	{\mbox{$>$}}

\newcommand{\cala}
        {\mathcal{A}}

\newcommand{\calb}
        {\mathcal{B}}

\newcommand{\calp}
        {\mathcal{P}}

\newcommand{\natns}
	{\mathbb{N}}

\newcommand{\procs}
	{\mathbb{P}}

\newcommand{\initial}
	{\textit{initial}}

\newcommand{\wf}
	{\textit{wf}}

\newcommand{\toinitial}
	{\textit{to\_initial}}

\newcommand{\upd}
	{\textit{upd}}

\newcommand{\frs}
	{\textit{frs}}

\newcommand{\brs}
	{\textit{brs}}

\newcommand{\xarrow}[2]
        {\, {\xrightarrow{#1}}_{#2} \,}

\newcommand{\arrow}[2]
        {\, {\auxarrow\limits^{#1}}_{#2} \,}
\newcommand{\auxarrow}
	{\mathop{\longrightarrow}}

\newcommand{\wauxarrow}
	{\mathop{\Longrightarrow}}

\newcommand{\nil}
	{\underline 0}

\newcommand{\eqdef}
	{\triangleq}

\newcommand{\parfun}
	{\rightharpoonup}
\newcommand{\sbis}[1]
	{\sim_{#1}}

\newcommand{\pco}[1]
	{\mathop{\Vert_{#1}}}

\newcommand{\lpar}
	{\rrfloor}

\newcommand{\rpar}
	{\llfloor}

\newcommand{\lplu}
	{\mathop{. \!\! +} \!}
\newcommand{\rplu}
	{\mathop{+ \!\! .} \hspace{-0.04cm}}
\newcommand{\size}
	{\textit{size}}

\newcommand{\act}
	{\textit{act}}

\newtheorem{new_theorem}
	{Theorem}[section]

\newtheorem{new_definition}
	[new_theorem]{Definition}

\newtheorem{new_remark}
	[new_theorem]{Remark}

\newtheorem{new_example}
	[new_theorem]{Example}

\newtheorem{new_lemma}
	[new_theorem]{Lemma}

\newtheorem{new_proposition}
	[new_theorem]{Proposition}

\newtheorem{new_corollary}
	[new_theorem]{Corollary}

\newenvironment{definition}
	{\begin{new_definition}\rm}
	{\end{new_definition}}

\newenvironment{example}
	{\begin{new_example}\rm}
	{\end{new_example}}

\newenvironment{lemma}
	{\begin{new_lemma}\rm}
	{\end{new_lemma}}

\newenvironment{proposition}
	{\begin{new_proposition}\rm}
	{\end{new_proposition}}

\newenvironment{theorem}
	{\begin{new_theorem}\rm}
	{\end{new_theorem}}

\newenvironment{corollary}
	{\begin{new_corollary}\rm}
	{\end{new_corollary}}

\begin{document}

\title{Expansion Laws for Forward-Reverse, Forward, and Reverse Bisimilarities via Proved Encodings}
\def\titlerunning{Expansion Laws for Forward-Reverse, Forward, and Reverse Bisimilarities via Proved
Encodings}

\author{Marco Bernardo \qquad Andrea Esposito \qquad Claudio A.\ Mezzina
\institute{Dipartimento di Scienze Pure e Applicate, Universit\`a di Urbino, Urbino, Italy}}
\def\authorrunning{M.\ Bernardo, A.\ Esposito \& C.A.\ Mezzina}

\maketitle

%%%%%%%%%%%%%%%%%%%%%%%%%%%%%%%%%%%%%%%%%%%%%%%%%%%%%%%%%%%%%%%%%
%                                                               %
%                                                               %
% Abstract                                                      %
%                                                               %
%                                                               %
%%%%%%%%%%%%%%%%%%%%%%%%%%%%%%%%%%%%%%%%%%%%%%%%%%%%%%%%%%%%%%%%%

\begin{abstract}
Reversible systems exhibit both forward computations and backward computations, where the aim of the latter
is to undo the effects of the former. Such systems can be compared via forward-reverse bisimilarity as well
as its two components, i.e., forward bisimilarity and reverse bisimilarity. The congruence, equational, and
logical properties of these equivalences have already been studied in the setting of sequential processes.
In this paper we address concurrent processes and investigate compositionality and axiomatizations of
forward bisimilarity, which is interleaving, and reverse and forward-reverse bisimilarities, which are truly
concurrent. To uniformly derive expansion laws for the three equivalences, we develop encodings based on the
proved trees approach of Degano \& Priami. In the case of reverse and forward-reverse bisimilarities, we
show that in the encoding every action prefix needs to be extended with the backward ready set of the
reached process.
\end{abstract}

%%%%%%%%%%%%%%%%%%%%%%%%%%%%%%%%%%%%%%%%%%%%%%%%%%%%%%%%%%%%%%%%%
%
%
\section{Introduction}
\label{sec:intro}
%
%
%%%%%%%%%%%%%%%%%%%%%%%%%%%%%%%%%%%%%%%%%%%%%%%%%%%%%%%%%%%%%%%%%

A reversible system features two directions of computation. The forward one coincides with the normal way of
computing. The backward one undoes the effects of the forward one so as to return to a consistent state,
i.e., a state that can be encountered while moving in the forward direction. Reversible computing has
attracted an increasing interest due to its applications in many areas, including low-power
computing~\cite{Lan61,Ben73}, program debugging~\cite{GLM14,LNPV18a}, robotics~\cite{LES18}, wireless
communications~\cite{SPP19}, fault-tolerant systems~\cite{DK05,VKH10,LLMSS13,VS18}, biochemical
modeling~\cite{PUY12,Pin17}, and parallel discrete-event simulation~\cite{PP14,SOJB18}.

Returning to a consistent state is not an easy task to accomplish in a concurrent system, because the undo
procedure necessarily starts from the last performed action and this may not be uniquely identifiable due to
concurrency. The usually adopted strategy is that an action can be undone provided that all the actions it
subsequently caused, if any, have been undone beforehand~\cite{DK04}. In this paper we focus on reversible
process calculi, for which there are two approaches -- later shown to be equivalent in~\cite{LMM21} --
\linebreak to keep track of executed actions and revert computations in a causality-consistent way.

The dynamic approach of~\cite{DK04,Kri12} yielded RCCS (R for reversible) and its mobile
variants~\cite{LMS10,CKV13}. RCCS is an extension of CCS~\cite{Mil89a} that uses stack-based memories
attached to processes so as to record executed actions and subprocesses discarded upon choices. A single
transition relation is defined, while actions are divided into forward and backward thereby resulting in
forward and backward transitions. This approach is adequate in the case of very expressive calculi as well
as programming languages.

The static approach of~\cite{PU07a} proposed a general method to reverse calculi, of which CCSK (K for keys)
and its quantitative variants~\cite{BM23a,BLMY24,BM23b,BM24b} are a result. The idea is to retain within the
process syntax all executed actions, which are suitably decorated, and all dynamic operators, which are thus
made static. A forward transition relation and a backward transition relation are defined separately. Their
labels are actions extended with communication keys so as to know, upon generating backward transitions,
which actions synchronized with each other. This approach is very handy to deal with basic process calculi.

A systematic study of compositionality and axiomatization of strong bisimilarity in reversible process
calculi has started in~\cite{BR23}, both for nondeterministic processes and for Markovian processes. Then
compositionality and axiomatization of weak bisimilarity as well as modal logic characterizations for strong
and weak bisimilarities have been investigated in~\cite{BE23a,BE23b} for the nondeterministic case. That
study compares the properties of forward-reverse bisimilarity $\sbis{\rm FRB}$~\cite{PU07a} with those of
its two components, i.e., forward bisimilarity $\sbis{\rm FB}$~\cite{Par81,Mil89a} and reverse bisimilarity
$\sbis{\rm RB}$. The reversible process calculus used in that study is minimal. Similar to~\cite{DMV90}, its
semantics relies on a single transition relation, where the distinction between going forward or backward in
the bisimulation game is made by matching outgoing or incoming transitions respectively. As a consequence,
similar to~\cite{BC94} executed actions can be decorated uniformly, without having to resort to external
stack-based memories~\cite{DK04} or communication keys associated with those actions~\cite{PU07a}.

	\begin{figure}\label{fig:expansion_law}

\centerline{\includegraphics{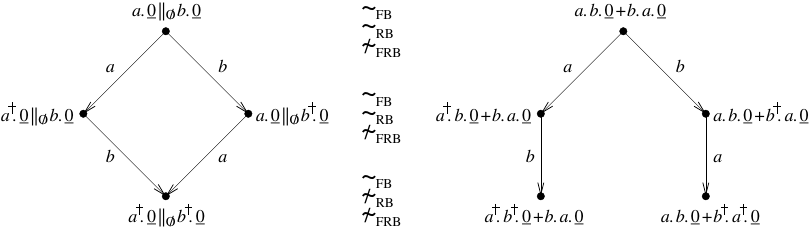}}
\caption{Forward, reverse, and forward-reverse bisimilarities at work: interleaving vs.\ true concurrency}

	\end{figure}

A substantial limitation of~\cite{BR23,BE23a,BE23b} is the absence of the parallel composition operator in
the calculus, motivated by the need of remaining neutral with respect to interleaving view vs.\ true
concurrency. Unlike forward bisimilarity, as noted in~\cite{PU07a} forward-reverse bisimilarity -- and also
reverse bisimilarity -- does not satisfy the expansion law of parallel composition into a nondeterministic
choice among all possible action sequencings. In Figure~\ref{fig:expansion_law} we depict two labeled
transition systems respectively representing a process that can perform action $a$ in parallel with action
$b$ ($a \, . \, \nil \pco{\emptyset} b \, . \, \nil$ using a CSP-like parallel composition~\cite{BHR84}) and
a process that can perform either $a$ followed by $b$ or $b$ followed by $a$ ($a \, . \, b \, . \, \nil + b
\, . \, a \, . \, \nil$ with $+$ denoting a CCS-like choice~\cite{Mil89a}), where $a \neq b$ and $\dag$
decorates executed actions.

The forward bisimulation game yields the usual interleaving setting in which the two top states are related,
the two pairs of corresponding intermediate states are related, and the three bottom states are related.
However, the three bottom states are no longer related if we play the reverse bisimulation game, as the
state on the left has two differently labeled incoming transitions while either state on the right has only
one. The remaining pairs of states are related by reverse bisimilarity as they have identically labeled
incoming transitions, whereas they are told apart by forward-reverse bisimilarity due to the failure of the
interplay between outgoing and incoming transitions matching. More precisely, any two corresponding
intermediate states are not forward-reverse bisimilar because their identically labeled outgoing transitions
reach the aforementioned inequivalent bottom states. In turn, the two initial states are not forward-reverse
bisimilar because their identically labeled outgoing transitions reach the aforementioned inequivalent
intermediate states. A new level of complexity thus arises from the introduction of parallel composition.

For the sake of completeness, we recall that an interleaving view can be restored by considering computation
paths (instead of states) like in the back-and-forth bisimilarity of~\cite{DMV90}. Besides causality, this
choice additionally preserves history, in the sense that backward moves are constrained to take place along
the path followed in the forward direction even in the presence of concurrency. For instance, in the labeled
transition system on the left, after performing $a$ and then $b$ it is not possible to undo $a$ before $b$
although there are no causality constraints between those two actions.

In this paper we add parallel composition and then extend the axiomatizations of the three strong
bisimilarities examined in~\cite{BR23} via expansion laws. The usual technique consists of introducing
normal forms, in which only action prefix and alternative composition occur, along with expansion laws,
through which occurrences of parallel composition are progressively eliminated. Although this originated in
the interleaving setting for forward-only calculi~\cite{HM85} to \emph{identify} processes such as $a \, .
\, \nil \pco{\emptyset} b \, . \, \nil$ and $a \, . \, b \, . \, \nil + b \, . \, a \, . \, \nil$, it was
later exploited also in the truly concurrent spectrum~\cite{GG01,Fec04} to \emph{distinguish} processes like
the aforementioned two. This requires an extension of the syntax that adds suitable discriminating
information within action prefixes. For example:

	\begin{itemize}

\item Causal bisimilarity~\cite{DD89,DD90} (corresponding to history-preserving bisimilarity~\cite{RT88}):
every action is enriched with the set of its causing actions, each of which is expressed as a numeric
backward pointer, so that the former process is expanded to $\lap a, \emptyset \rap \, . \, \lap b,
\emptyset \rap \, . \, \nil + \lap b, \emptyset \rap \, . \, \lap a, \emptyset \rap \, . \, \nil$ \linebreak
while the latter process becomes $\lap a, \emptyset \rap \, . \, \lap b, \{ 1 \} \rap \, . \, \nil + \lap b,
\emptyset \rap \, . \, \lap a, \{ 1 \} \rap \, . \, \nil$.

\item Location bisimilarity~\cite{BCHK94} (corresponding to local history-preserving
bisimilarity~\cite{Cas95}): every action is enriched with the name of the location in which it is executed,
so that the former process is expanded to $\lap a, l_{a} \rap \, . \, \lap b, l_{b} \rap \, . \, \nil + \lap
b, l_{b} \rap \, . \, \lap a, l_{a} \rap \, . \, \nil$ while the latter process becomes $\lap a, l_{a} \rap
\, . \, \lap b, l_{a} l_{b} \rap \, . \, \nil + \lap b, l_{b} \rap \, . \, \lap a, l_{b} l_{a} \rap \, . \,
\nil$.

\item Pomset bisimilarity~\cite{BC88a}: instead of a single action, a prefix may contain the combination of
several independent actions that are executed simultaneously, so that the former process is expanded to $a
\, . \, b \, . \, \nil + b \, . \, a \, . \, \nil + (a \pco{} b) \, . \, \nil$ while the latter process is
unchanged.

	\end{itemize}

A unifying framework for addressing both interleaving and truly concurrent semantics along with their
expansion laws was developed in~\cite{DP92}. The idea is to label every transition with a proof
term~\cite{BC88b,BC94}, which is an action preceded by the operators in the scope of which the action
occurs. The semantics of interest then drives an observation function that maps proof terms to the required
observations. In the interleaving case proof terms are reduced to the actions they contain, while in the
truly concurrent case they are transformed into actions extended with discriminating information as
exemplified above.

In this paper we apply the proved trees approach of~\cite{DP92} to develop expansion laws for forward,
reverse, and forward-reverse bisimilarities. This requires understanding which additional discriminating
information is needed inside prefixes for the last two equivalences. While this is rather straightforward
for the truly concurrent semantics recalled above -- the considered information is already present in the
original transition labels -- it is not obvious in our case because original transitions are labeled just
with actions. However, by looking at the three bottom states in Figure~\ref{fig:expansion_law}, one can
realize that they have different \emph{backward ready sets}, i.e., sets of actions labeling incoming
transitions: $\{ b, a \}, \{ b \}, \{ a \}$.

We show that backward ready sets indeed constitute the information that is necessary to add within action
prefixes for reverse and forward-reverse bisimilarities, by means of a suitable process encoding. Moreover,
we provide an adequate treatment of concurrent processes in which independent actions have been executed on
both sides of the parallel composition because, e.g., $a^{\dag} . \, \nil \pco{\emptyset} b^{\dag} . \,
\nil$ cannot be expanded to something like $a^{\dag} . \, b^{\dag} . \, \nil + b^{\dag} . \, a^{\dag} . \,
\nil$ in that only one branch of an alternative composition can be executed.

This paper is organized as follows. In Section~\ref{sec:from_seq_to_conc} we extend the syntax of the
reversible process calculus of~\cite{BR23} by adding a parallel composition operator, we reformulate its
operational semantics by following the proved trees approach of~\cite{DP92}, and we rephrase the definitions
of forward, reverse, and forward-reverse bisimilarities of~\cite{BR23}. In
Section~\ref{sec:obs_fun_proc_enc} we illustrate the next steps of the proved trees approach, i.e., the
definition of observation functions and process encodings. In Sections~\ref{sec:fb_exp_law}
and~\ref{sec:rb_frb_exp_law} \linebreak we respectively develop axioms for forward bisimilarity, including
an interleaving-style expansion law, and for reverse and forward-reverse bisimilarities, including expansion
laws based on extending action prefixes with backward ready sets. In Section~\ref{sec:concl} we provide some
concluding remarks.

%ZZZ
%The main purpose of axiomatizations of behavioral equivalences is to elucidate their fundamental laws, which
%would otherwise remain implicit in their definitions. As shown by Milner, expansion laws are also useful to
%relate sequential system specifications with their concurrent implementations. Although bisimulation
%equivalence checking is usually conducted via partition refinement algorithms like Kanellakis & Smolka's,
%axioms can be seen as rewriting rules and hence implemented in tools (mCRL2, Maude; future work). Given a
%process, computing backward ready sets once and for all via expansions would certainly be more convenient
%than going back and forth.

%%%%%%%%%%%%%%%%%%%%%%%%%%%%%%%%%%%%%%%%%%%%%%%%%%%%%%%%%%%%%%%%%
%
%
\section{From Sequential Reversible Processes to Concurrent Ones}
\label{sec:from_seq_to_conc}
%
%
%%%%%%%%%%%%%%%%%%%%%%%%%%%%%%%%%%%%%%%%%%%%%%%%%%%%%%%%%%%%%%%%%

Starting from the sequential reversible calculus considered in~\cite{BR23}, in this section we extend its
syntax with a parallel composition operator in the CSP style~\cite{BHR84} (Section~\ref{sec:syntax}) and its
semantics according to the proved trees approach~\cite{DP92} (Section~\ref{sec:semantics}). Then we rephrase
forward, reverse, and forward-reverse bisimilarities and show that they are congruences with respect to the
additional operator (Section~\ref{sec:bisim}).

%%%%%%%%%%%%%%%%%%%%%%%%%%%%%%%%%%%%%%%%%%%%%%%%%%%%%%%%%%%%%%%%%
%
\subsection{Syntax of Concurrent Reversible Processes}
\label{sec:syntax}
%
%%%%%%%%%%%%%%%%%%%%%%%%%%%%%%%%%%%%%%%%%%%%%%%%%%%%%%%%%%%%%%%%%

Given a countable set $A$ of actions including an unobservable action denoted by~$\tau$, the syntax of
concurrent reversible processes extends the one in~\cite{BR23} as follows:
\cws{0}{P \:\: ::= \:\: \nil \mid a \, . \, P \mid a^{\dag} . \, P \mid P + P \mid \color{blue}{P \pco{L}
P}}
where $a \in A$, $\dag$ decorates executed actions, $L \subseteq A \setminus \{ \tau \}$, and:

	\begin{itemize}

\item $\nil$ is the terminated process.

\item $a \, . \, P$ is a process that can execute action $a$ and whose forward continuation is $P$.

\item $a^{\dag} . \, P$ is a process that executed action $a$ and whose forward continuation is inside $P$,
which can undo action $a$ after all executed actions within $P$ have been undone.

\item $P_{1} + P_{2}$ expresses a nondeterministic choice between $P_{1}$ and $P_{2}$ as far as neither has
executed any action yet, otherwise only the one that was selected in the past can move.

\item $P_{1} \pco{L} P_{2}$ expresses the parallel composition of $P_{1}$ and $P_{2}$, which proceed
independently of each other on actions in $\bar{L} = A \setminus L$ while they have to synchronize on every
action in $L$.

	\end{itemize}

As in~\cite{BR23} we can characterize some important classes of processes via as many predicates. Firstly,
we define \emph{initial} processes, in which all actions are unexecuted and hence no $\dag$-decoration
appears:
\cws{0}{\begin{array}{rcl}
\initial(\nil) & & \\
\initial(a \, . \, P) & \!\!\! \textrm{if} \!\!\! & \initial(P) \\
\initial(P_{1} + P_{2}) & \!\!\! \textrm{if} \!\!\! & \initial(P_{1}) \land \initial(P_{2}) \\
\initial(P_{1} \pco{L} P_{2}) & \!\!\! \textrm{if} \!\!\! & \initial(P_{1}) \land \initial(P_{2}) \\
\end{array}}
\indent
Secondly, we define \emph{well-formed} processes, whose set we denote by $\calp$, in which both unexecuted
and executed actions can occur in certain circumstances:
\cws{0}{\begin{array}{rcl}
\wf(\nil) & & \\
\wf(a \, . \, P) & \!\!\! \textrm{if} \!\!\! & \initial(P) \\
\wf(a^{\dag} . \, P) & \!\!\! \textrm{if} \!\!\! & \wf(P) \\
\wf(P_{1} + P_{2}) & \!\!\! \textrm{if} \!\!\! & (\wf(P_{1}) \land \initial(P_{2})) \lor (\initial(P_{1})
\land \wf(P_{2})) \\
\wf(P_{1} \pco{L} P_{2}) & \!\!\! \textrm{if} \!\!\! & \wf(P_{1}) \land \wf(P_{2}) \\
\end{array}}
Well formedness not only imposes that every unexecuted action is followed by an initial process, but also
that in every alternative composition at least one subprocess is initial. Multiple paths arise in the
presence of both alternative ($+$) and parallel ($\pco{L}$) compositions. However, at each occurrence of the
former, only the subprocess chosen for execution can move. Although not selected, the other subprocess is
kept as an initial subprocess within the overall process in the same way as executed actions are kept inside
the syntax~\cite{BC94,PU07a}, so as to support reversibility. For example, in $a^{\dag} . \, b \, . \, \nil
+ c \, . \, d \, . \, \nil$ the subprocess $c \, . \, d \, . \, \nil$ cannot move as $a$ was selected in the
choice between $a$ and $c$.

It is worth noting that:

	\begin{itemize}

\item $\nil$ is both initial and well-formed.

\item Any initial process is well-formed too.

\item $\calp$ also contains processes that are not initial like, e.g., $a^{\dag} . \, b \, . \, \nil$, which
can either do $b$ or undo $a$.

\item In $\calp$ the relative positions of already executed actions and actions to be executed matter.
Precisely, an action of the former kind can never occur after one of the latter kind. For instance,
$a^{\dag} . \, b \, . \, \nil \in \calp$ whereas $b \, . \, a^{\dag} . \, \nil \notin \calp$.

\item In $\calp$ the subprocesses of an alternative composition can be both initial, but cannot be both
non-initial. As an example, $a \, . \, \nil + b \, . \, \nil \in \calp$ whilst $a^{\dag} . \, \nil +
b^{\dag} . \, \nil \notin \calp$.

	\end{itemize}

%%%%%%%%%%%%%%%%%%%%%%%%%%%%%%%%%%%%%%%%%%%%%%%%%%%%%%%%%%%%%%%%%
%
\subsection{Proved Operational Semantics}
\label{sec:semantics}
%
%%%%%%%%%%%%%%%%%%%%%%%%%%%%%%%%%%%%%%%%%%%%%%%%%%%%%%%%%%%%%%%%%

According to~\cite{PU07a}, in the semantic rules dynamic operators such as action prefix and alternative
composition have to be made static, so as to retain within the syntax all the information needed to enable
reversibility. Unlike~\cite{PU07a}, we do not generate a forward transition relation and a backward one, but
a single transition relation that, like in~\cite{DMV90}, we deem to be symmetric in order to enforce the
\emph{loop property}~\cite{DK04}: every executed action can be undone and every undone action can be redone.
In our setting, \linebreak a backward transition from $P'$ to~$P$ is subsumed by the corresponding forward
transition $t$ from $P$ to $P'$. \linebreak As we will see in the definition of behavioral equivalences,
like in~\cite{DMV90} we view $t$ as an \emph{outgoing} transition of $P$ when going forward, while we view
$t$ as an \emph{incoming} transition of $P'$ when going backward.

Unlike~\cite{BR23}, as a first step based on~\cite{DP92} towards the derivation of expansion laws for
parallel composition we provide a very concrete semantics in which every transition is labeled with a
\emph{proof term}~\cite{BC88b,BC94}. This is an action preceded by the sequence of operator symbols in the
scope of which the action occurs. In the case of a binary operator, the corresponding symbol also specifies
whether the action occurs to the left or to the right. The syntax that we adopt for the set $\Theta$ of
proof terms is the following:
\cws{0}{\theta \:\: ::= \:\: a \mid . \theta \mid \lplu \theta \mid \rplu \theta \mid \,\, \lpar \theta \mid
\rpar \theta \mid \langle \theta, \theta \rangle}
\indent
The proved semantic rules in Table~\ref{tab:proved_semantics} extend the ones in~\cite{BR23} and generate
the proved labeled transition system $(\calp, \Theta, \! \arrow{}{} \!)$ where $\! \arrow{}{} \! \subseteq
\calp \times \Theta \times \calp$ is the proved transition relation. We denote by $\procs \subsetneq \calp$
\linebreak the set of processes that are \emph{reachable} from an initial one via $\! \arrow{}{} \!$. Not
all well-formed processes are reachable; for example, $a^{\dag} . \, \nil \pco{\{ a \}} \nil$ is not
reachable from $a \, . \, \nil \pco{\{ a \}} \nil$ as action $a$ on the left cannot synchronize with any
action on the right. We indicate with $\procs_{\rm init}$ the set of initial processes in $\procs$.

	\begin{table}[t]

\[\begin{array}{|ll|}
\hline
\inferrule*[left=(Act$_{\rm f}$)]{\initial(P)}{a \, . \, P \arrow{a}{} a^{\dag} . \, P} &
\inferrule*[left=(Act$_{\rm p}$)]{P \arrow{\theta}{} P'}{a^{\dag} . \, P \arrow{{\color{magenta}{.}}
\theta}{} a^{\dag} . \, P'} \\[0.2cm]
\inferrule*[left=(Cho$_{\rm l}$)]{P_{1} \arrow{\theta}{} P'_{1} \quad \initial(P_{2})}{P_{1} + P_{2}
\arrow{{\color{magenta}{\lplu}} \, \theta}{} P'_{1} + P_{2}} \quad &
\inferrule*[left=(Cho$_{\rm r}$)]{P_{2} \arrow{\theta}{} P'_{2} \quad \initial(P_{1})}{P_{1} + P_{2}
\arrow{{\color{magenta}{\rplu}} \, \theta}{} P_{1} + P'_{2}} \\[0.2cm]
\inferrule*[left=(Par$_{\rm l}$)]{P_{1} \arrow{\theta}{} P'_{1} \quad \act(\theta) \notin L}{P_{1} \pco{L}
P_{2} \arrow{{\color{magenta}{\lpar}} \theta}{} P'_{1} \pco{L} P_{2}} \quad &
\inferrule*[left=(Par$_{\rm r}$)]{P_{2} \arrow{\theta}{} P'_{2} \quad \act(\theta) \notin L}{P_{1} \pco{L}
P_{2} \arrow{{\color{magenta}{\rpar}} \theta}{} P_{1} \pco{L} P'_{2}} \\[0.2cm]
& \hspace*{-4.4cm} \inferrule*[left=(Syn)]{P_{1} \arrow{\theta_{1}}{} P'_{1} \quad P_{2}
\arrow{\theta_{2}}{} P'_{2} \quad \act(\theta_{1}) = \act(\theta_{2}) \in L}{P_{1} \pco{L} P_{2}
\xarrow{{\color{magenta}{\langle}} \theta_{1} \! {\color{magenta}{,}} \theta_{2}
{\color{magenta}{\rangle}}}{} P'_{1} \pco{L} P'_{2}} \\
\hline
\end{array}\]

\caption{Proved operational semantic rules for concurrent reversible processes}
\label{tab:proved_semantics}

	\end{table}

The first rule for action prefix (\textsc{Act}$_{\rm f}$ where f stands for forward) applies only if $P$ is
initial and retains the executed action in the target process of the generated forward transition by
decorating the action itself with $\dag$. The second rule (\textsc{Act}$_{\rm p}$ where p stands for
propagation) propagates actions of inner initial subprocesses by putting a dot before them in the label for
each outer executed action prefix.

In both rules for alternative composition (\textsc{Cho}$_{\rm l}$ and \textsc{Cho}$_{\rm r}$ where l stands
for left and r~stands for right), the subprocess that has not been selected for execution is retained as an
initial subprocess in the target process of the generated transition. When both subprocesses are initial,
both rules for alternative composition are applicable, otherwise only one of them can be applied and in that
case it is the non-initial subprocess that can move, because the other one has been discarded at the moment
of the selection.

The rules for parallel composition make use of partial function $\act : \Theta \parfun A$ to extract the
action from a proof term $\theta$. The function is defined by induction on the syntactical structure of
$\theta$ as follows:
\cws{0}{\begin{array}{rcl}
\act(a) & \!\!\! = \!\!\! & a \\
\act(. \theta') \: = \: \act(\lplu \theta') \: = \: \act(\hspace{-0.04cm} \rplu \theta') \: = \: \act(\lpar
\theta') \: = \: \act(\rpar \theta') & \!\!\! = \!\!\! & \act(\theta') \\
\act(\langle \theta_{1}, \theta_{2} \rangle) & \!\!\! = \!\!\! & \act(\theta_{1}) \quad \textrm{if
$\act(\theta_{1}) = \act(\theta_{2})$} \\
\end{array}}
In the first two rules (\textsc{Par}$_{\rm l}$ and \textsc{Par}$_{\rm r}$), a single subprocess proceeds by
performing an action not belonging to $L$. In the third rule (\textsc{Syn}), both subprocesses synchronize
on an action in $L$.

Every process may have several outgoing transitions and, if it is not initial, has at least one incoming
transition. Due to the decoration of executed actions inside the process syntax, over the set $\procs_{\rm
seq}$ of \emph{sequential} processes -- in which there are no occurrences of parallel composition -- every
non-initial process has exactly one incoming transition, the underlying labeled transition systems turn out
to be trees, and well formedness coincides with reachability~\cite{BR23}.

	\begin{example}\label{ex:semantics}

The proved labeled transition systems generated by the rules in Table~\ref{tab:proved_semantics} for the two
initial sequential processes $a \, . \, \nil + a \, . \, \nil$ and $a \, . \, \nil$ are depicted below:
\\[0.1cm]
\centerline{\includegraphics{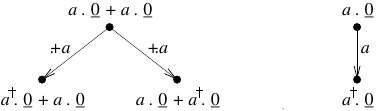}}
In the case of a forward-only process calculus, a single $a$-transition would be generated from $a \, . \,
\nil + a \, . \, \nil$ to $\nil$ due to the absence of action decorations within processes.
\fullbox

	\end{example}

%%%%%%%%%%%%%%%%%%%%%%%%%%%%%%%%%%%%%%%%%%%%%%%%%%%%%%%%%%%%%%%%%
%
\subsection{Forward, Reverse, and Forward-Reverse Bisimilarities}
\label{sec:bisim}
%
%%%%%%%%%%%%%%%%%%%%%%%%%%%%%%%%%%%%%%%%%%%%%%%%%%%%%%%%%%%%%%%%%

We rephrase the definitions given in~\cite{BR23} of forward bisimilarity~\cite{Par81,Mil89a} (only
\emph{outgoing} transitions), reverse bisimilarity (only \emph{incoming} transitions), and forward-reverse
bisimilarity~\cite{PU07a} (both kinds of transitions) because transition labels now are proof terms. Since
we focus on the actions contained in those terms, the distinguishing power of the three equivalences does
not change with respect to~\cite{BR23}.

	\begin{definition}\label{def:forward_bisim}

We say that $P_{1}, P_{2} \in \procs$ are \emph{forward bisimilar}, written $P_{1} \sbis{\rm FB} P_{2}$, iff
$(P_{1}, P_{2}) \in \calb$ for some forward bisimulation $\calb$. A symmetric relation $\calb$ over~$\procs$
is a \emph{forward bisimulation} iff, whenever $(P_{1}, P_{2}) \in \calb$, then:

		\begin{itemize}

\item For each $P_{1} \arrow{\theta_{1}}{} P'_{1}$ there exists $P_{2} \arrow{\theta_{2}}{} P'_{2}$ such
that $\act(\theta_{1}) = \act(\theta_{2})$ and $(P'_{1}, P'_{2}) \in \calb$.
\fullbox

		\end{itemize}

	\end{definition}

	\begin{definition}\label{def:reverse_bisim}

We say that $P_{1}, P_{2} \in \procs$ are \emph{reverse bisimilar}, written $P_{1} \sbis{\rm RB} P_{2}$, iff
$(P_{1}, P_{2}) \in \calb$ for \linebreak some reverse bisimulation $\calb$. A symmetric relation $\calb$
over $\procs$ is a \emph{reverse bisimulation} iff, whenever $(P_{1}, P_{2}) \in \calb$, then:

		\begin{itemize}

\item For each $P'_{1} \arrow{\theta_{1}}{} P_{1}$ there exists $P'_{2} \arrow{\theta_{2}}{} P_{2}$ such
that $\act(\theta_{1}) = \act(\theta_{2})$ and $(P'_{1}, P'_{2}) \in \calb$.
\fullbox

		\end{itemize}

	\end{definition}

\pagebreak

	\begin{definition}\label{def:forward_reverse_bisim}

We say that $P_{1}, P_{2} \in \procs$ are \emph{forward-reverse bisimilar}, written $P_{1} \sbis{\rm FRB}
P_{2}$, iff $(P_{1}, P_{2}) \in \calb$ for some forward-reverse bisimulation $\calb$. A symmetric relation
$\calb$ over $\procs$ is a \emph{forward-reverse bisimulation} iff, whenever $(P_{1}, P_{2}) \in \calb$,
then:

		\begin{itemize}

\item For each $P_{1} \arrow{\theta_{1}}{} P'_{1}$ there exists $P_{2} \arrow{\theta_{2}}{} P'_{2}$ such
that $\act(\theta_{1}) = \act(\theta_{2})$ and $(P'_{1}, P'_{2}) \in \calb$.

\item For each $P'_{1} \arrow{\theta_{1}}{} P_{1}$ there exists $P'_{2} \arrow{\theta_{2}}{} P_{2}$ such
that $\act(\theta_{1}) = \act(\theta_{2})$ and $(P'_{1}, P'_{2}) \in \calb$.
\fullbox

		\end{itemize}

	\end{definition}

	\begin{example}\label{ex:bisim}

The two initial processes considered in Example~\ref{ex:semantics} are identified by all the three
equivalences. This is witnessed by any bisimulation that contains the pairs $(a \, . \, \nil + a \, . \,
\nil, a \, . \, \nil)$, $(a^{\dag} . \, \nil + a \, . \, \nil, a^{\dag} . \, \nil)$, and $(a \, . \, \nil +
a^{\dag} . \, \nil, a^{\dag} . \, \nil)$.
\fullbox

	\end{example}

As observed in~\cite{BR23}, $\sbis{\rm FB}$ is not a congruence with respect to alternative composition,
e.g.:
\cws{0}{\begin{array}{rclcrcl}
a^{\dag} . \, b \, . \, \nil & \!\!\! \sbis{\rm FB} \!\!\! & b \, . \, \nil & \textrm{but} &
a^{\dag} . \, b \, . \, \nil + c \, . \, \nil & \!\!\! \not\sbis{\rm FB} \!\!\! & b \, . \, \nil + c \, . \,
\nil \\
\end{array}}
because in $a^{\dag} . \, b \, . \, \nil + c \, . \, \nil$ action $c$ is disabled by virtue of the already
executed action $a^{\dag}$, while in $b \, . \, \nil + c \, . \, \nil$ action $c$ is enabled as there are no
past actions preventing it from occurring. This problem, which does not show up for $\sbis{\rm RB}$ and
$\sbis{\rm FRB}$ because they cannot identify an initial process with a non-initial one, led in~\cite{BR23}
to the following variant of~$\sbis{\rm FB}$ that is sensitive to the presence of the past.

	\begin{definition}\label{def:ps_forward_bisim}

We say that $P_{1}, P_{2} \in \procs$ are \emph{past-sensitive forward bisimilar}, written $P_{1} \sbis{\rm
FB:ps} P_{2}$, iff $(P_{1}, P_{2}) \in \calb$ for some past-sensitive forward bisimulation~$\calb$. A
relation $\calb$ over $\procs$ is a \emph{past-sensitive forward bisimulation} iff it is a forward
bisimulation where $\initial(P_{1}) \Longleftrightarrow \initial(P_{2})$ for all $(P_{1}, P_{2}) \in \calb$.~\fullbox

	\end{definition}

Since $\sbis{\rm FB:ps}$ is sensitive to the presence of the past, we have that $a^{\dag} . \, b \, . \,
\nil \: \not\sbis{\rm FB:ps} \: b \, . \, \nil$, but it is still possible to identify non-initial processes
having a different past like, e.g., $a_{1}^{\dag} \, . \, P$ and $a_{2}^{\dag} \, . \, P$. It holds that
$\sbis{\rm FRB} \; \subsetneq \; \sbis{\rm FB:ps} \cap \sbis{\rm RB}$, with $\sbis{\rm FRB} \, = \,
\sbis{\rm FB:ps}$ over initial processes as well as $\sbis{\rm FB:ps}$ and $\sbis{\rm RB}$ being \linebreak
incomparable because, e.g., for $a_{1} \neq a_{2}$:
\cws{0}{\begin{array}{rcl}
a_{1}^{\dag} \, . \, P \, \sbis{\rm FB:ps} \, a_{2}^{\dag} \, . \, P & \textrm{but} & a_{1}^{\dag} \, . \,
P \, \not\sbis{\rm RB} \, a_{2}^{\dag} \, . \, P \\
a_{1} \, . \, P \, \sbis{\rm RB} \, a_{2} \, . \, P & \textrm{but} & a_{1} \, . \, P \, \not\sbis{\rm FB:ps}
\, a_{2} \, . \, P \\
\end{array}}
\indent
It is easy to establish two necessary conditions for the considered bisimilarities. Following the
terminology of~\cite{OH86,BKO88}, the two conditions respectively make use of the forward ready set in the
forward direction and the backward ready set in the backward direction; the latter condition will be
exploited when developing the expansion laws for $\sbis{\rm RB}$ and $\sbis{\rm FRB}$. We proceed by
induction on the syntactical structure of $P \in \procs$ to define its \emph{forward ready set} $\frs(P)
\subseteq A$, i.e., the set of actions that $P$ can immediately execute (labels of its outgoing
transitions), as well as its \emph{backward ready set} $\brs(P) \subseteq A$, i.e., the set of actions whose
execution led to $P$ (labels of its incoming transitions):
\cws{0}{\begin{array}{rclcrcl}
\frs(\nil) & \!\!\! = \!\!\! & \emptyset & &
\brs(\nil) & \!\!\! = \!\!\! & \emptyset \\
\frs(a \, . \, P') & \!\!\! = \!\!\! & \{ a \} & &
\brs(a \, . \, P') & \!\!\! = \!\!\! & \emptyset \\
\frs(a^{\dag} . \, P') & \!\!\! = \!\!\! & \frs(P') & &
\brs(a^{\dag} . \, P') & \!\!\! = \!\!\! & \left\{ \begin{array}{ll}
\{ a \} & \textrm{if $\initial(P')$} \\
\brs(P') & \textrm{if $\lnot\initial(P')$} \\
\end{array} \right. \\
\end{array}}
\cws{0}{\begin{array}{rcl}
\frs(P_{1} + P_{2}) & \!\!\! = \!\!\!\! & \left\{ \begin{array}{ll}
\frs(P_{1}) \cup \frs(P_{2}) & \textrm{if $\initial(P_{1}) \land \initial(P_{2})$} \\
\frs(P_{1}) & \textrm{if $\lnot\initial(P_{1}) \land \initial(P_{2})$} \\
\frs(P_{2}) & \textrm{if $\initial(P_{1}) \land \lnot\initial(P_{2})$} \\
\end{array} \right. \\
\brs(P_{1} + P_{2}) & \!\!\! = \!\!\!\! & \left\{ \begin{array}{ll}
\emptyset & \textrm{if $\initial(P_{1}) \land \initial(P_{2})$} \\
\brs(P_{1}) & \textrm{if $\lnot\initial(P_{1}) \land \initial(P_{2})$} \\
\brs(P_{2}) & \textrm{if $\initial(P_{1}) \land \lnot\initial(P_{2})$} \\
\end{array} \right. \\
\frs(P_{1} \pco{L} P_{2}) & \!\!\! = \!\!\! & (\frs(P_{1}) \cap \bar{L}) \cup (\frs(P_{2}) \cap \bar{L})
\cup (\frs(P_{1}) \cap \frs(P_{2}) \cap L) \\
\brs(P_{1} \pco{L} P_{2}) & \!\!\! = \!\!\! & (\brs(P_{1}) \cap \bar{L}) \cup (\brs(P_{2}) \cap \bar{L})
\cup (\brs(P_{1}) \cap \brs(P_{2}) \cap L) \\
\end{array}}

	\begin{proposition}\label{prop:bisim_nec_cond}

Let $P_{1}, P_{2} \in \procs$. Then:

		\begin{enumerate}

\item If $P_{1} \sbis{} P_{2}$ for $\sbis{} \, \in \{ \sbis{\rm FB}, \sbis{\rm FB:ps}, \sbis{\rm FRB} \}$,
then $\frs(P_{1}) = \frs(P_{2})$.

\item If $P_{1} \sbis{} P_{2}$ for $\sbis{} \, \in \{ \sbis{\rm RB}, \sbis{\rm FRB} \}$, then $\brs(P_{1}) =
\brs(P_{2})$.
\fullbox

		\end{enumerate}

	\end{proposition}

In~\cite{BR23} it has been shown that all these four bisimilarities are congruences with respect to action
prefix, while only $\sbis{\rm FB:ps}$, $\sbis{\rm RB}$, and $\sbis{\rm FRB}$ are congruences with respect to
alternative composition too, with $\sbis{\rm FB:ps}$ being the coarsest congruence with respect to $+$
contained in~$\sbis{\rm FB}$. Sound and ground-complete equational characterizations have also been provided
for the three congruences. Here we show that all these bisimilarities are congruences with respect to the
newly added operator, i.e., parallel composition.

	\begin{theorem}\label{thm:congr_par_comp}

Let $\sbis{} \, \in \{ \sbis{\rm FB}, \sbis{\rm FB:ps}, \sbis{\rm RB}, \sbis{\rm FRB} \}$ and $P_{1}, P_{2}
\in \procs$. If $P_{1} \sbis{} P_{2}$ then $P_{1} \pco{L} P \sbis{} P_{2} \pco{L} P$ and $P \pco{L} P_{1}
\sbis{} P \pco{L} P_{2}$ for all $P \in \procs$ and $L \subseteq A \setminus \{ \tau \}$ such that $P_{1}
\pco{L} P, P_{2} \pco{L} P, P \pco{L} P_{1}, P \pco{L} P_{2} \in \procs$.
\fullbox

	\end{theorem}

%%%%%%%%%%%%%%%%%%%%%%%%%%%%%%%%%%%%%%%%%%%%%%%%%%%%%%%%%%%%%%%%%
%
%
\section{Observation Functions and Process Encodings for Expansion Laws}
\label{sec:obs_fun_proc_enc}
%
%
%%%%%%%%%%%%%%%%%%%%%%%%%%%%%%%%%%%%%%%%%%%%%%%%%%%%%%%%%%%%%%%%%

Among the most important axioms there are \emph{expansion laws}, which are useful to relate sequential
specifications of systems with their concurrent implementations~\cite{Mil89a}. In the interleaving setting
they can be obtained quite naturally, whereas this is not the case under true concurrency. Thanks to the
proved operational semantics in Table~\ref{tab:proved_semantics}, we can uniformly derive expansion laws for
the interleaving bisimulation congruence $\sbis{\rm FB:ps}$ and the two truly concurrent bisimulation
congruences $\sbis{\rm RB}$ and $\sbis{\rm FRB}$ by following the proved trees approach of~\cite{DP92}.

All we have to do is the introduction of three \emph{observation functions} $\ell_{\rm F}$, $\ell_{\rm R}$,
and $\ell_{\rm FR}$ that respectively transform the proof terms labeling all proved transitions into
suitable observations according to $\sbis{\rm FB:ps}$, $\sbis{\rm RB}$, and $\sbis{\rm FRB}$. In addition to
a specific proof term $\theta$, as shown in~\cite{DP92} each such function, say $\ell$, may depend on other
possible parameters in the proved labeled transition system generated by the semantic rules in
Table~\ref{tab:proved_semantics}. Moreover, it must preserve actions, i.e., if $\ell(\theta_{1}) =
\ell(\theta_{2})$ then $\act(\theta_{1}) = \act(\theta_{2})$.

Based on the corresponding $\ell$, from each of the three aforementioned congruences we can thus derive a
bisimilarity in which, whenever $(P_{1}, P_{2}) \in \calb$, the forward clause requires that: \\
\centerline{for each $P_{1} \xarrow{\ell(\theta_{1})}{} P'_{1}$ there exists $P_{2}
\xarrow{\ell(\theta_{2})}{} P'_{2}$ such that $\ell(\theta_{1}) = \ell(\theta_{2})$ and $(P'_{1}, P'_{2})
\in \calb$}
while the backward clause requires that: \\
\centerline{for each $P'_{1} \xarrow{\ell(\theta_{1})}{} P_{1}$ there exists $P'_{2}
\xarrow{\ell(\theta_{2})}{} P_{2}$ such that $\ell(\theta_{1}) = \ell(\theta_{2})$ and $(P'_{1}, P'_{2}) \in
\calb$}
We indicate with $\sbis{{\rm FB:ps}:\ell_{\rm F}}$, $\sbis{{\rm RB}:\ell_{\rm R}}$, and $\sbis{{\rm
FRB}:\ell_{\rm FR}}$ the three resulting bisimilarities.\label{pnt:ell_bisim}

To derive the corresponding expansion laws, the final step -- left implicit in~\cite{DP92}, see, e.g., the
forthcoming Definitions~\ref{def:brs_enc_proc} and~\ref{def:brs_enc_proc_el} -- consists of lifting $\ell$
to processes so as to encode observations within action prefixes. For a process $P \in \procs_{\rm seq}$,
the idea is to proceed by induction on the syntactical structure of $P$ \linebreak as follows, where $\sigma
\in \Theta_{\rm seq}^{*}$ for $\Theta_{\rm seq} = \{ ., \lplu \hspace{0.03cm}, \rplu \hspace{0.04cm} \}$:
\cws{0}{\begin{array}{rcl}
\ell^{\sigma}(\nil) & \!\!\! = \!\!\! & \nil \\
\ell^{\sigma}(a \, . \, P') & \!\!\! = \!\!\! & \ell(\sigma a) \, . \, \ell^{\sigma
{\color{magenta}{.}}}(P') \\
\ell^{\sigma}(a^{\dag} . \, P') & \!\!\! = \!\!\! & \ell(\sigma a)^{\dag} . \, \ell^{\sigma
{\color{magenta}{.}}}(P') \\
\ell^{\sigma}(P_{1} + P_{2}) & \!\!\! = \!\!\! & \ell^{{{\color{magenta}{\lplu}} \, \sigma}}(P_{1}) +
\ell^{{{\color{magenta}{\rplu}} \, \sigma}}(P_{2}) \\
\end{array}}
Every sequential process $P$ will thus be encoded as $\ell^{\varepsilon}(P)$, so for example $a \, . \, b \,
. \, \nil + b \, . \, a \, . \, \nil$ will become:
\cws{0}{\ell^{\lplu}(a \, . \, b \, . \, \nil) + \ell^{\rplu}(b \, . \, a \, . \, \nil) \, = \, \ell(\lplu
a) \, . \, \ell^{\lplu .}(b \, . \, \nil) + \ell(\rplu b) \, . \, \ell^{\rplu .}(a \, . \, \nil) \, = \,
\ell(\lplu a) \, . \, \ell(\lplu . b) \, . \, \nil + \ell(\rplu b) \, . \, \ell(\rplu . a) \, . \, \nil}
\indent
Then, given two initial sequential processes expressed as follows due to the commutativity and associativity
of alternative composition (where any summation over an empty index set is $\nil$):
\cws{0}{P_{1} \: = \: \sum\limits_{i \in I_{1}} \ell(\theta_{1, i}) \, . \, P_{1, i} \quad \textrm{and}
\quad P_{2} \: = \: \sum\limits_{i \in I_{2}} \ell(\theta_{2, i}) \, . \, P_{2, i}}
the idea is to encode their parallel composition via the following expansion law (where $\nil \pco{L} \nil$
yields $\nil$):
\cws{0}{\begin{array}{rcl}
P_{1} \pco{L} P_{2} & \!\!\! = \!\!\! & \sum\limits_{i \in I_{1}, \act(\theta_{1, i}) \notin L}
\ell({\color{magenta}{\lpar}} \theta_{1, i}) \, . \, (P_{1, i} \pco{L} P_{2}) + \sum\limits_{i \in I_{2},
\act(\theta_{2, i}) \notin L} \ell({\color{magenta}{\rpar}} \theta_{2, i}) \, . \, (P_{1} \pco{L} P_{2, i})
\; + \\
& & \sum\limits_{i \in I_{1}, \act(\theta_{1, i}) \in L} \; \sum\limits_{j \in I_{2}, \act(\theta_{2, j}) =
\act(\theta_{1, i})} \ell({\color{magenta}{\langle}} \theta_{1, i} \! {\color{magenta}{,}} \theta_{2, j}
{\color{magenta}{\rangle}}) \, . \, (P_{1, i} \pco{L} P_{2, j}) \\
\end{array}}
For instance, $a \, . \, \nil \pco{\emptyset} b \, . \, \nil$, represented as $\ell(a) \, . \, \nil
\pco{\emptyset} \ell(b) \, . \, \nil$, will be expanded as follows:
\cws{0}{\ell(\lpar_{\emptyset} a) \, . \, (\nil \pco{\emptyset} \ell(b) \, . \, \nil) +
\ell(\rpar_{\emptyset} b) \, . \, (\ell(a) \, . \, \nil \pco{\emptyset} \nil) \: = \: \ell(\lpar_{\emptyset}
a) \, . \, \ell(\rpar_{\emptyset} b) \, . \, \nil + \ell(\rpar_{\emptyset} b) \, . \, \ell(\lpar_{\emptyset}
a) \, . \, \nil}
where, compared to the encoding of $a \, . \, b \, . \, \nil + b \, . \, a \, . \, \nil$, in general
$\ell(\lplu a) \neq \ell(\lpar_{\emptyset} a) \neq \ell(\rplu . a)$ and $\ell(\lplu . b) \neq
\ell(\rpar_{\emptyset} b) \neq \ell(\rplu b)$. The expansion laws for the cases in which the two sequential
processes are not both initial -- which are specific to reversible processes and hence not addressed
in~\cite{DP92} -- are derived similarly. We will see that care must be taken when both processes are
non-initial because for example $a^{\dag} . \, \nil \pco{\emptyset} b^{\dag} . \, \nil$ cannot be expanded
to $\ell(\lpar a)^{\dag} . \, \ell(\rpar b)^{\dag} . \, \nil + \ell(\rpar b)^{\dag} . \, \ell(\lpar
a)^{\dag} . \, \nil$ as the latter is not even well-formed due to the presence of executed actions on both
sides of the alternative composition.

In the next two sections we will investigate how to define the three observation functions $\ell_{\rm F}$,
$\ell_{\rm R}$, and $\ell_{\rm FR}$ in such a way that the three equivalences $\sbis{{\rm FB:ps}:\ell_{\rm
F}}$, $\sbis{{\rm RB}:\ell_{\rm R}}$, and $\sbis{{\rm FRB}:\ell_{\rm FR}}$ respectively coincide with the
three congruences $\sbis{\rm FB:ps}$, $\sbis{\rm RB}$, and $\sbis{\rm FRB}$. As far as true concurrency is
concerned, we point out that the observation functions developed in~\cite{DP92} for causal semantics and
location semantics were inspired by additional information already present in the labels of the original
semantics, i.e., backward pointers sets~\cite{DD89} and localities~\cite{BCHK94} respectively. In our case,
instead, the original semantics in Table~\ref{tab:proved_semantics} features labels that are essentially
actions, hence for reverse and forward-reverse bisimilarities we have to find out the additional information
necessary to discriminate, e.g., the processes associated with the three bottom states in
Figure~\ref{fig:expansion_law}.

%%%%%%%%%%%%%%%%%%%%%%%%%%%%%%%%%%%%%%%%%%%%%%%%%%%%%%%%%%%%%%%%%
%
%
\section{Axioms and Expansion Law for $\sbis{\rm FB:ps}$}
\label{sec:fb_exp_law}
%
%
%%%%%%%%%%%%%%%%%%%%%%%%%%%%%%%%%%%%%%%%%%%%%%%%%%%%%%%%%%%%%%%%%

In this section we provide a sound and ground-complete axiomatization of forward bisimilarity over
concurrent reversible processes. As already mentioned, this behavioral equivalence complies with the
interleaving view of concurrency. Therefore, we can exploit the same observation function for interleaving
semantics used in~\cite{DP92}, which we express as $\ell_{\rm F}(\theta) = \act(\theta)$ and immediately
implies that $\sbis{{\rm FB:ps}:\ell_{\rm F}}$ coincides with $\sbis{\rm FB:ps}$. Moreover, no additional
information has to be inserted into action prefixes, i.e., the lifting to processes of the observation
function is accomplished via the identity function.

	\begin{table}[t]

\[\begin{array}{|lrcll|}
\hline
& & & & \\[-0.4cm]
(\cala_{{\rm F}, 1}) & (P + Q) + R & \!\!\! = \!\!\! & P + (Q + R) & \hspace*{-4.2cm} \textrm{where at least
two among $P$, $Q$, $R$ are initial} \\
(\cala_{{\rm F}, 2}) & P + Q & \!\!\! = \!\!\! & Q + P & \hspace*{-4.2cm} \textrm{where at least one between
$P$ and $Q$ is initial} \\
(\cala_{{\rm F}, 3}) & P + \nil & \!\!\! = \!\!\! & P & \\
(\cala_{{\rm F}, 4}) & P + P & \!\!\! = \!\!\! & P & \hspace*{-4.2cm} \textrm{where $\initial(P)$} \\
(\cala_{{\rm F}, 5}) & a^{\dag} \, . \, P & \!\!\! = \!\!\! & b^{\dag} \, . \, P & \hspace*{-4.2cm}
\textrm{if $\initial(P)$} \\
(\cala_{{\rm F}, 6}) & a^{\dag} . \, P & \!\!\! = \!\!\! & P & \hspace*{-4.2cm} \textrm{if
$\lnot\initial(P)$} \\
(\cala_{{\rm F}, 7}) & P + Q & \!\!\! = \!\!\! & P & \hspace*{-4.2cm} \textrm{if $\lnot\initial(P)$, where
$\initial(Q)$} \\
(\cala_{{\rm F}, 8}) & P_{1} \pco{L} P_{2} & \!\!\! = \!\!\! & [a^{\dag} .] \left( \sum\limits_{i \in I_{1},
a_{1, i} \notin L} a_{1, i} \, . \, (P_{1, i} \pco{L} P'_{2}) \; + \right. & \\[0.4cm]
& & & \hspace*{1.05cm} \sum\limits_{i \in I_{2}, a_{2, i} \notin L} a_{2, i} \, . \, (P'_{1} \pco{L} P_{2,
i}) \; + & \\[-0.2cm]
& & & \hspace*{1.0cm} \left. \sum\limits_{i \in I_{1}, a_{1, i} \in L} \; \sum\limits_{j \in I_{2}, a_{2,
j} = a_{1, i}} a_{1, i} \, . \, (P_{1, i} \pco{L} P_{2, j}) \right) & \\[0.4cm]
& & & & \hspace*{-9.2cm} \textrm{with $P_{k} = [a_{k}^{\dag} .] P'_{k}, \; P'_{k} = \sum\limits_{i \in
I_{k}} a_{k, i} \, . \, P_{k, i}$ in F-nf for $k \in \{ 1, 2 \}$ and $a^{\dag}$ present iff so is
$a_{1}^{\dag}$ or $a_{2}^{\dag}$} \\
\hline
\end{array}\]

\caption{Axioms characterizing $\sbis{\rm FB:ps}$ over concurrent reversible processes}
\label{tab:fb_axioms}

	\end{table}

The set $\cala_{\rm F}$ of axioms for $\sbis{\rm FB:ps}$ is shown in Table~\ref{tab:fb_axioms}
(where-clauses are related to $\procs$-membership). All the axioms apart from the last one come
from~\cite{BR23}, where an axiomatization was developed over sequential reversible processes.
Axioms~$\cala_{{\rm F}, 1}$ to~$\cala_{{\rm F}, 4}$ -- associativity, commutativity, neutral element, and
idempotency of alternative composition -- coincide with those for forward-only processes~\cite{HM85}.
Axioms~$\cala_{{\rm F}, 5}$ and~$\cala_{{\rm F}, 6}$ together establish that the presence of the past cannot
be ignored, but the specific past can be neglected when moving only forward. Likewise, axiom~$\cala_{{\rm
F}, 7}$ states that a previously non-selected alternative process can be discarded when moving only forward;
note that it does not subsume axioms~$\cala_{{\rm F}, 3}$ and~$\cala_{{\rm F}, 4}$ because $P$ has to be
non-initial.

Since due to axioms~$\cala_{{\rm F}, 5}$ and~$\cala_{{\rm F}, 6}$ we only need to remember whether some
action has been executed in the past, axiom~$\cala_{{\rm F}, 8}$ is the only expansion law needed for
$\sbis{\rm FB:ps}$. Notation $[a^{\dag} .]$ stands for the possible presence of an executed action prefix,
with $a^{\dag}$ being present at the beginning of the expansion iff at least one of $a_{1}^{\dag}$ and
$a_{2}^{\dag}$ is present at the beginning of $P_{1}$ and $P_{2}$ respectively. In $P_{1}$ and $P_{2}$, as
well as on the righthand side of the expansion, summations are allowed by axioms~$\cala_{{\rm F}, 1}$
and~$\cala_{{\rm F}, 2}$ and represent $\nil$ when their index sets are empty (so that $\cala_{\rm F} \vdash
\nil \pco{L} \nil = \nil + \nil + \nil = \nil$ due to axiom~$\cala_{{\rm F}, 3}$, substitutivity with
respect to alternative composition, and transitivity).

The deduction system based on $\cala_{\rm F}$, whose deducibility relation we denote by $\, \vdash$,
includes axioms and inference rules expressing reflexivity, symmetry, and transitivity (because $\sbis{\rm
FB:ps}$ is an equivalence relation) as well as substitutivity with respect to the operators of the
considered calculus (because $\sbis{\rm FB:ps}$ is a congruence with respect to all of those operators).
Following~\cite{HM85}, to show the soundness and ground-completeness of $\cala_{\rm F}$ for $\sbis{\rm
FB:ps}$ we introduce a suitable normal form to which every process can be reduced. The only operators that
can occur in such a normal form are action prefix and alternative composition, hence all processes in normal
form are sequential.

	\begin{definition}\label{def:f_normal_form}

We say that $P \in \procs$ is in \emph{forward normal form}, written \emph{F-nf}, iff it is equal to
$[b^{\dag} .] \sum_{i \in I} a_{i} \, . \, P_{i}$ where the executed action prefix $b^{\dag} . \, \_$ is
optional, $I$ is a finite index set (with the summation being $\nil$ \linebreak when $I = \emptyset$), and
each $P_{i}$ is initial and in F-nf.
\fullbox

	\end{definition}

	\begin{lemma}\label{lem:reduc_fnf}

For all (initial) $P \in \procs$ there exists (an initial) $Q \in \procs$ in F-nf such that $\cala_{\rm F}
\vdash P = Q$.
\fullbox

	\end{lemma}

	\begin{theorem}\label{thm:fb_axioms}

Let $P_{1}, P_{2} \in \procs$. Then $P_{1} \sbis{\rm FB:ps} P_{2}$ iff $\cala_{\rm F} \vdash P_{1} = P_{2}$.
\fullbox

	\end{theorem}

%%%%%%%%%%%%%%%%%%%%%%%%%%%%%%%%%%%%%%%%%%%%%%%%%%%%%%%%%%%%%%%%%
%
%
\section{Axioms and Expansion Laws for $\sbis{\rm RB}$ and $\sbis{\rm FRB}$}
\label{sec:rb_frb_exp_law}
%
%
%%%%%%%%%%%%%%%%%%%%%%%%%%%%%%%%%%%%%%%%%%%%%%%%%%%%%%%%%%%%%%%%%

In this section we address the axiomatization of reverse and forward-reverse bisimilarities over concurrent
reversible processes. Since these behavioral equivalences are truly concurrent, we have to provide process
encodings that insert suitable additional discriminating information into action prefixes. We show that this
information is the same for both semantics and is constituted by backward ready sets. Precisely, for every
proved transition $P \arrow{\theta}{} P'$, we let $\ell_{\rm R}(\theta)_{P'} = \ell_{\rm FR}(\theta)_{P'} =
\lap \act(\theta), \brs(P') \rap \eqdef \ell_{\rm brs}(\theta)_{P'}$, where in the observation function we
have indicated its primary argument $\theta$ in parentheses and its secondary argument $P'$ as a subscript
(see Section~\ref{sec:obs_fun_proc_enc} for the possibility of additional parameters like $P'$).

By virtue of Proposition~\ref{prop:bisim_nec_cond}$(2)$, the distinguishing power of $\sbis{\rm RB}$ and
$\sbis{\rm FRB}$ does not change if, in the related definitions of bisimulation, we additionally require
that $\brs(P_{1}) = \brs(P_{2})$ for all $(P_{1}, P_{2}) \in \calb$. As a consequence, it is straightforward
to realize that $\sbis{{\rm RB}:\ell_{\rm brs}}$ and $\sbis{{\rm FRB}:\ell_{\rm brs}}$ (see
page~\pageref{pnt:ell_bisim}) respectively coincide with $\sbis{\rm RB}$ and $\sbis{\rm FRB}$ over~$\procs$.
Moreover, $\sbis{{\rm RB}:\ell_{\rm brs}}$ and $\sbis{{\rm FRB}:\ell_{\rm brs}}$ also apply to the encoding
target $\procs_{\rm brs}$, i.e., the set of processes obtained from $\procs_{\rm seq}$ by extending every
action prefix with a subset of $A$.

	\begin{table}[t]

\[\begin{array}{|ll|}
\hline
\inferrule*[left=(Act$_{\rm brs, f}$)]{\initial(U)}{\lap a, \beth \rap \, . \, U \xarrow{a,
{\color{cyan}{\beth}}}{\rm brs} \lap a^{\dag}, \beth \rap \, . \, U} &
\inferrule*[left=(Act$_{\rm brs, p}$)]{U \xarrow{\theta, {\color{cyan}{\daleth}}}{\rm brs} U'}{\lap
a^{\dag}, \beth \rap \, . \, U \xarrow{. \theta, {\color{cyan}{\daleth}}}{\rm brs} \lap a^{\dag}, \beth \rap
\, . \, U'} \\[0.2cm]
\inferrule*[left=(Cho$_{\rm brs, l}$)]{U_{1} \xarrow{\theta, {\color{cyan}{\beth}}}{\rm brs} U'_{1} \quad
\initial(U_{2})}{U_{1} + U_{2} \xarrow{\lplu \theta, {\color{cyan}{\beth}}}{\rm brs} U'_{1} + U_{2}} \quad &
\inferrule*[left=(Cho$_{\rm brs, r}$)]{U_{2} \xarrow{\theta, {\color{cyan}{\beth}}}{\rm brs} U'_{2} \quad
\initial(U_{1})}{U_{1} + U_{2} \xarrow{\rplu \theta, {\color{cyan}{\beth}}}{\rm brs} U_{1} + U'_{2}} \\
\hline
\end{array}\]

\caption{Proved operational semantic rules for $\procs_{\rm brs}$ ($\beth, \daleth \in 2^{A}$)}
\label{tab:brs_semantics}

	\end{table}

The syntax of $\procs_{\rm brs}$ processes is defined as follows where $\beth \in 2^{A}$:
\cws{0}{U \:\: ::= \:\: \nil \mid \lap a, {\color{cyan}{\beth}} \rap \, . \, U \mid \lap a^{\dag},
{\color{cyan}{\beth}} \rap \, . \, U \mid U + U}
The proved operational semantic rules for $\procs_{\rm brs}$ shown in Table~\ref{tab:brs_semantics} generate
the proved labeled transition system $(\procs_{\rm brs}, \Theta \times 2^{A}, \! \arrow{}{\rm brs} \!)$.
With respect to those in Table~\ref{tab:proved_semantics}, the rules in Table~\ref{tab:brs_semantics}
additionally label the produced transitions with the action sets occurring in the action prefixes inside the
source processes. Given a symmetric relation $\calb$ over $\procs_{\rm brs}$, whenever $(U_{1}, U_{2}) \in
\calb$ the forward clause of $\sbis{{\rm FRB}:\ell_{\rm brs}}$ can be rephrased as: \\
\centerline{for each $U_{1} \xarrow{\theta_{1}, \beth}{\rm brs} U'_{1}$ there exists $U_{2}
\xarrow{\theta_{2}, \beth}{\rm brs} U'_{2}$ such that $\act(\theta_{1}) = \act(\theta_{2})$ and $(U'_{1},
U'_{2}) \in \calb$}
while the backward clauses of $\sbis{{\rm RB}:\ell_{\rm brs}}$ and $\sbis{{\rm FRB}:\ell_{\rm brs}}$ can be
rephrased as: \\
\centerline{for each $U'_{1} \xarrow{\theta_{1}, \beth}{\rm brs} U_{1}$ there exists $U'_{2}
\xarrow{\theta_{2}, \beth}{\rm brs} U_{2}$ such that $\act(\theta_{1}) = \act(\theta_{2})$ and $(U'_{1},
U'_{2}) \in \calb$}

Following the proved trees approach as described in Section~\ref{sec:obs_fun_proc_enc}, we now lift
$\ell_{\rm brs}$ so as to encode $\procs$ into $\procs_{\rm brs}$. The objective is to extend each action
prefix with the backward ready set of the reached process. While in the case of processes in $\procs_{\rm
seq}$ it is just a matter of extending any action prefix with a singleton containing the action itself,
backward ready sets may contain several actions when handling processes not in $\procs_{\rm seq}$. To
account for this, the lifting of $\ell_{\rm brs}$ has to make use of a secondary argument, which we
call environment process and will be written as a subscript by analogy with the secondary argument of the
observation function.

The environment process is progressively updated as prefixes are turned into pairs so as to represent the
process reached so far, i.e., the process yielding the backward ready set. The environment process $E$
for $P$ embodies $P$, in the sense that it is initially $P$ and then its forward behavior is updated upon
each action prefix extension by decorating the action as executed, where the action is located within $E$ by
a proof term. To correctly handle the extension of already executed prefixes, (part of) $E$ has to be
brought back by replacing $P$ inside $E$ with the process $\toinitial(P)$ obtained from $P$ by removing all
$\dag$-decorations. Function $\toinitial : \procs \rightarrow \procs_{\rm init}$ is defined by induction on
the syntactical structure of $P \in \procs$ as follows:
\cws{0}{\begin{array}{rcll}
\toinitial(P) & \!\!\! = \!\!\! & P & \textrm{if $\initial(P)$} \\
\toinitial(a^{\dag} . \, P') & \!\!\! = \!\!\! & a \, . \, \toinitial(P') & \\
\toinitial(P_{1} + P_{2}) & \!\!\! = \!\!\! & \toinitial(P_{1}) + \toinitial(P_{2}) & \textrm{if
$\lnot\initial(P_{1}) \lor \lnot\initial(P_{2})$} \\
\toinitial(P_{1} \pco{L} P_{2}) & \!\!\! = \!\!\! & \toinitial(P_{1}) \pco{L} \toinitial(P_{2}) & \textrm{if
$\lnot\initial(P_{1}) \lor \lnot\initial(P_{2})$} \\
\end{array}}
\indent
In Definitions~\ref{def:brs_enc_proc} and~\ref{def:brs_enc_proc_el} we develop the lifting of $\ell_{\rm
brs}$ and denote by $\widetilde{P}$ the result of its application. We recall that $\ell_{\rm
brs}(\theta)_{P'} = \lap \act(\theta), \brs(P') \rap$ and we let $\ell_{\rm brs}(\theta)^{\dag}_{P'} = \lap
\act(\theta)^{\dag}, \brs(P') \rap$. We further recall that $\Theta_{\rm seq} = \{ ., \lplu \hspace{0.03cm},
\rplu \hspace{0.04cm} \}$.

	\begin{definition}\label{def:brs_enc_proc}

Let $P \in \procs$, $E \in \procs$ be such that $P$ is a subprocess of $E$, and $\ddot{E}$ be obtained from
$E$ \linebreak by replacing $P$ with $\toinitial(P)$. The \emph{$\ell_{\rm brs}$-encoding} of $P$ is
$\widetilde{P} = \ell_{\rm brs}^{\varepsilon}(P)_{P}$, where $\ell_{\rm brs}^{\sigma} : \procs \times \procs
\rightarrow \procs_{\rm brs}$ \linebreak for $\sigma \in \Theta_{\rm seq}^{*}$ is defined by induction on
the syntactical structure of its primary argument $P \in \procs$ \linebreak (its secondary argument is $E
\in \procs$) as follows:
\cws{0}{\begin{array}{rcl}
\ell_{\rm brs}^{\sigma}(\nil)_{E} & \!\!\! = \!\!\! & \nil \\
\ell_{\rm brs}^{\sigma}(a \, . \, P')_{E} & \!\!\! = \!\!\! & \ell_{\rm brs}(\sigma a)_{\upd(E, \sigma a)}
\, . \, \ell_{\rm brs}^{\sigma .}(P')_{\upd(E, \sigma a)} \\
\ell_{\rm brs}^{\sigma}(a^{\dag} . \, P')_{E} & \!\!\! = \!\!\! & \ell_{\rm brs}(\sigma
a)^{\dag}_{\upd(\ddot{E}, \sigma a)} . \, \ell_{\rm brs}^{\sigma .}(P')_{E} \\
\ell_{\rm brs}^{\sigma}(P_{1} + P_{2})_{E} & \!\!\! = \!\!\! & \ell_{\rm brs}^{\sigma \! \lplu}(P_{1})_{E} +
\ell_{\rm brs}^{\sigma \! \rplu}(P_{2})_{E} \\
\ell_{\rm brs}^{\sigma}(P_{1} \pco{L} P_{2})_{E} & \!\!\! = \!\!\! & e\ell_{\rm
brs}^{\sigma}(\widetilde{P}_{1}, \widetilde{P}_{2}, L)_{E} \\
\end{array}}
where function $e\ell_{\rm brs}^{\sigma}$ will be defined later on while function $\upd : \procs \times
\Theta \rightarrow \procs$ is defined by induction on the syntactical structural of its first argument $E
\in \procs$ as follows:
\cws{10}{\begin{array}{rcl}
\upd(\nil, \theta) & \!\!\! = \!\!\! & \nil \\
\upd(a \, . \, E', \theta) & \!\!\! = \!\!\! & \left\{ \begin{array}{ll}
a^{{\color{purple}{\dag}}} . \, E' & \textrm{if $\theta = a$} \\
a \, . \, E' & \textrm{otherwise} \\
\end{array} \right. \\
\upd(a^{\dag} . \, E', \theta) & \!\!\! = \!\!\! & \left\{ \begin{array}{ll}
a^{\dag} . \, \upd(E', \theta') & \textrm{if $\theta = . \theta'$} \\
a^{\dag} . \, E' & \textrm{otherwise} \\
\end{array} \right. \\
\upd(E_{1} + E_{2}, \theta) & \!\!\! = \!\!\! & \left\{ \begin{array}{ll}
\upd(E_{1}, \theta') + E_{2} & \textrm{if $\theta = \lplu \theta'$} \\
E_{1} + \upd(E_{2}, \theta') & \textrm{if $\theta = \rplu \theta'$} \\
E_{1} + E_{2} & \textrm{otherwise} \\
\end{array} \right. \\
\upd(E_{1} \pco{L} E_{2}, \theta) & \!\!\! = \!\!\! & \left\{ \begin{array}{ll}
\upd(E_{1}, \theta') \pco{L} E_{2} & \textrm{if $\theta = \, \lpar \theta'$} \\
E_{1} \pco{L} \upd(E_{2}, \theta') & \textrm{if $\theta = \rpar \theta'$} \\
\upd(E_{1}, \theta_{1}) \pco{L} \upd(E_{2}, \theta_{2}) & \textrm{if $\theta = \langle \theta_{1},
\theta_{2} \rangle$} \\
E_{1} \pco{L} E_{2} & \textrm{otherwise} \\
\end{array} \right. \\
\end{array}}
\fullbox

	\end{definition}

	\begin{example}\label{ex:brs_enc_seq_proc}

Encoding sequential processes (for them function $e\ell_{\rm brs}^{\sigma}$ does not come into play):

		\begin{itemize}

\item Let $P$ be the initial sequential process $a \, . \, b \, . \, \nil + b \, . \, a \, . \, \nil$. Then:
\cws{8}{\hspace*{-1.0cm}\begin{array}{rcl}
\widetilde{P} \: = \: \ell_{\rm brs}^{\varepsilon}(P)_{P} & \!\!\! = \!\!\! & \ell_{\rm brs}^{\lplu}(a \, .
\, b \, . \, \nil)_{a \, . \, b \, . \, \nil + b \, . \, a \, . \, \nil} + \ell_{\rm brs}^{\rplu}(b \, . \,
a \, . \, \nil)_{a \, . \, b \, . \, \nil + b \, . \, a \, . \, \nil} \\
& \!\!\! = \!\!\! & \ell_{\rm brs}(\lplu a)_{a^{\dag} . \, b \, . \, \nil + b \, . \, a \, . \, \nil} \, .
\, \ell_{\rm brs}^{\lplu .}(b \, . \, \nil)_{a^{\dag} . \, b \, . \, \nil + b \, . \, a \, . \, \nil} \; +
\\
& & \ell_{\rm brs}(\rplu b)_{a \, . \, b \, . \, \nil + b^{\dag} . \, a \, . \, \nil} \, . \, \ell_{\rm
brs}^{\rplu .}(a \, . \, \nil)_{a \, . \, b \, . \, \nil + b^{\dag} . \, a \, . \, \nil} \\
& \!\!\! = \!\!\! & \lap a, \{ a \} \rap \, . \, \ell_{\rm brs}(\lplu . b)_{a^{\dag} . \, b^{\dag} . \, \nil
+ b \, . \, a \, . \, \nil} \, . \, \ell_{\rm brs}^{\lplu . .}(\nil)_{a^{\dag} . \, b^{\dag} . \, \nil + b
\, . \, a \, . \, \nil} \; + \\
& & \lap b, \{ b \} \rap \, . \, \ell_{\rm brs}(\rplu . a)_{a \, . \, b \, . \, \nil + b^{\dag} . \,
a^{\dag} . \, \nil} \, . \, \ell_{\rm brs}^{\rplu . .}(\nil)_{a \, . \, b \, . \, \nil + b^{\dag} . \,
a^{\dag} . \, \nil} \\
& \!\!\! = \!\!\! & \lap a, \{ a \} \rap \, . \, \lap b, \{ b \} \rap \, . \, \nil + \lap b, \{ b \} \rap \,
. \, \lap a, \{ a \} \rap \, . \, \nil \\
\end{array}}

\item Let $P$ be the non-initial sequential process $a^{\dag} . \, b^{\dag} . \, \nil$. Then:
\cws{0}{\hspace*{-1.0cm}\begin{array}{rclcl}
\widetilde{P} \: = \: \ell_{\rm brs}^{\varepsilon}(P)_{P} & \!\!\! = \!\!\! & \ell_{\rm brs}(a)_{a^{\dag} .
\, b \, . \, \nil}^{\dag} \, . \, \ell_{\rm brs}^{.}(b^{\dag} . \, \nil)_{a^{\dag} . \, b^{\dag} . \, \nil}
\hspace{0.1cm} = & & \\
& \!\!\! = \!\!\! & \lap a^{\dag}, \{ a \} \rap \, . \, \ell_{\rm brs}(. b)_{a^{\dag} . \, b^{\dag} . \,
\nil}^{\dag} \, . \, \ell_{\rm brs}^{. .}(\nil)_{a^{\dag} . \, b^{\dag} . \, \nil} & \!\!\! = \!\!\! & \lap
a^{\dag}, \{ a \} \rap \, . \, \lap b^{\dag}, \{ b \} \rap \, . \, \nil \\
\end{array}}
Definition~\ref{def:brs_enc_proc} yields $a \, . \, b \, . \, \nil$ as $\ddot{P}$ after the second $=$
(before it, $P$ is a subprocess of the environment $P$) and $a^{\dag} . \, b \, . \, \nil$ as $\ddot{P}$
after the third $=$ (before it, $b^{\dag} . \, \nil$ is a subprocess of the environment $P$).
\fullbox

		\end{itemize}

	\end{example}

While for sequential processes the backward ready set added to every action prefix is a singleton containing
the action itself, this is no longer the case when dealing with processes of the form $P_{1} \pco{L} P_{2}$.
We thus complete the encoding by providing the definition of $e\ell_{\rm brs}^{\sigma}$. When $P_{1}$ and
$P_{2}$ are not both initial, in the expansion we have to reconstruct all possible alternative action
sequencings that have not been undertaken -- which yield as many initial subprocesses -- because under the
forward-reverse semantics each of them could be selected after a rollback. In the subcase in which both
$P_{1}$ and $P_{2}$ are non-initial and their executed actions are not in $L$ -- e.g., $a^{\dag} . \, \nil
\pco{\emptyset} b^{\dag} . \, \nil$ -- care must be taken because executed actions cannot appear on both
sides of an alternative composition -- e.g., the expansion cannot be $a^{\dag} . \, b^{\dag} . \, \nil +
b^{\dag} . \, a^{\dag} . \, \nil$ in that not well-formed. To overcome this, based on a total order
$\le_{\dag}$ over $\Theta$ induced by the trace of actions executed so far, the expansion builds the
corresponding sequencing of already executed actions plus all the aforementioned unexecuted action
sequencings -- e.g., $a^{\dag} . \, b^{\dag} . \, \nil + b \, . \, a \, . \, \nil$ or $b^{\dag} . \,
a^{\dag} . \, \nil + a \, . \, b \, . \, \nil$ depending on whether $\lpar a \le_{\dag} \! \rpar b$ or
$\rpar b \le_{\dag} \, \lpar a$ respectively.

	\begin{definition}\label{def:brs_enc_proc_el}

Let $P_{1}, P_{2} \in \procs$, $L \subseteq A \setminus \{ \tau \}$, $E_{1}, E_{2}, E \in \procs$ be such
that $P_{1} \pco{L} P_{2}, E_{1} \pco{L} E_{2} \in \procs$, $P_{1}$ is a subprocess of~$E_{1}$, $P_{2}$ is a
subprocess of $E_{2}$, and $E_{1} \pco{L} E_{2}$ is a subprocess of $E$. Then $e\ell_{\rm brs}^{\sigma} :
\procs_{\rm brs} \times \procs_{\rm brs} \times 2^{A \setminus \{ \tau \}} \times \procs \rightarrow
\procs_{\rm brs}$ for $\sigma \in \Theta_{\rm seq}^{*}$ is inductively defined as follows, where square
brackets enclose optional subprocesses as already done in Section~\ref{sec:fb_exp_law} and every summation
over an empty index set is taken to be $\nil$ (and for simplicity is omitted within a choice unless all
alternative subprocesses inside that choice are $\nil$, in which case the whole choice boils down to
$\nil$):

		\begin{itemize}

\item If $\widetilde{P}_{1}$ and $\widetilde{P}_{2}$ are both initial, say $\widetilde{P}_{k} = \sum_{i \in
I_{k}} \ell_{\rm brs}(\theta_{k, i})_{\upd(P_{k}, \theta_{k, i})} \, . \, \widetilde{P}_{k, i}$ for $k \in
\{ 1, 2 \}$, let $e\ell_{\rm brs}^{\sigma}(\widetilde{P}_{1}, \widetilde{P}_{2}, L)_{E}$
\cws{0}{\hspace*{-1.4cm}\begin{array}{rl}
= \!\!\!\!\!\! & \sum\limits_{i \in I_{1}, \act(\theta_{1, i}) \notin L} \ell_{\rm brs}(\sigma
{\color{magenta}{\lpar}} \theta_{1, i})_{\upd(E, \sigma \lpar \theta_{1, i})} \, . \, e\ell_{\rm
brs}^{\sigma}(\widetilde{P}_{1, i}, \widetilde{P}_{2}, L)_{\upd(E, \sigma \lpar \theta_{1, i})} \; + \\
& \sum\limits_{i \in I_{2}, \act(\theta_{2, i}) \notin L} \ell_{\rm brs}(\sigma {\color{magenta}{\rpar}}
\theta_{2, i})_{\upd(E, \sigma \rpar \theta_{2, i})} \, . \, e\ell_{\rm brs}^{\sigma}(\widetilde{P}_{1},
\widetilde{P}_{2, i}, L)_{\upd(E, \sigma \rpar \theta_{2, i})} \; + \\
& \sum\limits_{i \in I_{1}, \act(\theta_{1, i}) \in L} \; \sum\limits_{j \in I_{2}, \act(\theta_{2, j}) =
\act(\theta_{1, i})} \hspace{-0.5cm} \ell_{\rm brs}(\sigma {\color{magenta}{\langle}} \theta_{1, i}
{\color{magenta}{,}} \theta_{2, j} {\color{magenta}{\rangle}})_{\upd(E, \sigma \langle \theta_{1, i},
\theta_{2, j} \rangle)} \, . \, e\ell_{\rm brs}^{\sigma}(\widetilde{P}_{1, i}, \widetilde{P}_{2, j},
L))_{\upd(E, \sigma \langle \theta_{1, i}, \theta_{2, j} \rangle)} \\
\end{array}}
where each of the three summation-shaped subprocesses on the right is an initial process.

\item If $\widetilde{P}_{1}$ is not initial while $\widetilde{P}_{2}$ is initial, say $\widetilde{P}_{1} =
\ell_{\rm brs}(\theta_{1})^{\dag}_{\upd(\toinitial(P_{1}), \theta_{1})} \, . \, \widetilde{P}'_{1} \, [+ \,
\widetilde{P}''_{1}]$ where $\act(\theta_{1}) \notin L$ \linebreak and $\widetilde{P}''_{1}$ is initial, say
$\widetilde{P}''_{1} = \sum_{i \in I_{1}} \ell_{\rm brs}(\theta_{1, i})_{\upd(P''_{1}, \theta_{1, i})} \, .
\, \widetilde{P}''_{1, i}$, and $\widetilde{P}_{2} = \sum_{i \in I_{2}} \ell_{\rm brs}(\theta_{2,
i})_{\upd(P_{2}, \theta_{2, i})} \, . \, \widetilde{P}_{2, i}$, \linebreak for $\ddot{E}$ obtained from $E$
by replacing $P_{1}$ with $\toinitial(P_{1})$ let $e\ell_{\rm brs}^{\sigma}(\widetilde{P}_{1},
\widetilde{P}_{2}, L)_{E}$
\cws{0}{\hspace*{-1.0cm}\begin{array}{rl}
= \!\!\! & \ell_{\rm brs}(\sigma \lpar \theta_{1})^{\color{magenta}{\dag}}_{\upd(\ddot{E}, \sigma \lpar
\theta_{1})} \, . \, e\ell_{\rm brs}^{\sigma}(\widetilde{P}'_{1}, \widetilde{P}_{2}, L)_{E} \; + \\
& [\sum_{i \in I_{1}, \act(\theta_{1, i}) \notin L} \ell_{\rm brs}(\sigma \lpar \theta_{1,
i})_{\upd(\ddot{E}, \sigma \lpar \theta_{1, i})} \, . \, e\ell_{\rm brs}^{\sigma}(\widetilde{P}''_{1, i},
\widetilde{P}_{2}, L)_{\upd(\ddot{E}, \sigma \lpar \theta_{1, i})} \; +] \\
& \sum_{i \in I_{2}, \act(\theta_{2, i}) \notin L} \ell_{\rm brs}(\sigma \rpar \theta_{2,
i})_{\upd(\ddot{E}, \sigma \rpar \theta_{2, i})} \, . \, e\ell_{\rm
brs}^{\sigma}(\toinitial(\widetilde{P}_{1}), \widetilde{P}_{2, i}, L)_{\upd(\ddot{E}, \sigma \rpar
\theta_{2, i})} \; + \\
& [\sum\limits_{i \in I_{1}, \act(\theta_{1, i}) \in L} \; \sum\limits_{j \in I_{2}, \act(\theta_{2, j}) =
\act(\theta_{1, i})} \hspace{-0.5cm} \ell_{\rm brs}(\sigma \langle \theta_{1, i}, \theta_{2, j}
\rangle)_{\upd(\ddot{E}, \sigma \langle \theta_{1, i}, \theta_{2, j} \rangle)} \, . \, e\ell_{\rm
brs}^{\sigma}(\widetilde{P}''_{1, i}, \widetilde{P}_{2, j}, L))_{\upd(\ddot{E}, \sigma \langle \theta_{1,
i}, \theta_{2, j} \rangle)}] \\
\end{array}}
where each of the last three summation-shaped subprocesses on the right is an initial process needed by the
forward-reverse semantics, with the presence of the first one and the third one \linebreak depending on the
presence of $\widetilde{P}''_{1}$.

\item The case in which $\widetilde{P}_{1}$ is initial while $\widetilde{P}_{2}$ is not initial is like the
previous one.

\item If $\widetilde{P}_{1}$ and $\widetilde{P}_{2}$ are both non-initial, say $\widetilde{P}_{k} =
\ell_{\rm brs}(\theta_{k})^{\dag}_{\upd(\toinitial(P_{k}), \theta_{k})} \, . \, \widetilde{P}'_{k} \, [+ \,
\widetilde{P}''_{k}]$ where $\widetilde{P}''_{k}$ is initial, say $\widetilde{P}''_{k} = \sum_{i \in I_{k}}
\ell_{\rm brs}(\theta_{k, i})_{\upd(P''_{k}, \theta_{k, i})} \, . \, \widetilde{P}''_{k, i}$, for $k \in \{
1, 2 \}$, for $\ddot{E}$ obtained from $E$ by replacing each $P_{k}$ with $\toinitial(P_{k})$ there are
three subcases:

			\begin{itemize}

\item If $\act(\theta_{1}) \notin L \land (\act(\theta_{2}) \in L \lor \sigma \lpar \theta_{1} \le_{\dag}
\sigma \rpar \theta_{2})$, let $e\ell_{\rm brs}^{\sigma}(\widetilde{P}_{1}, \widetilde{P}_{2}, L)_{E}$
\cws{0}{\hspace*{-1.5cm}\begin{array}{rl}
= \!\!\! & \ell_{\rm brs}(\sigma \lpar \theta_{1})^{\color{magenta}{\dag}}_{\upd(\ddot{E}, \sigma \lpar
\theta_{1})} \, . \, e\ell_{\rm brs}^{\sigma}(\widetilde{P}'_{1}, \widetilde{P}_{2}, L)_{E} \; + \\
& [\ell_{\rm brs}(\sigma \rpar \theta_{2})_{\upd(\ddot{E}, \sigma \rpar \theta_{2})} \, . \, e\ell_{\rm
brs}^{\sigma}(\toinitial(\widetilde{P}_{1}), \toinitial(\widetilde{P}'_{2}), L)_{\upd(\ddot{E}, \sigma \rpar
\theta_{2})} \; +] \\
& [\sum\limits_{i \in I_{1}, \act(\theta_{1, i}) \notin L} \ell_{\rm brs}(\sigma \lpar \theta_{1,
i})_{\upd(\ddot{E}, \sigma \lpar \theta_{1, i})} \, . \, e\ell_{\rm brs}^{\sigma}(\widetilde{P}''_{1, i},
\toinitial(\widetilde{P}_{2}), L)_{\upd(\ddot{E}, \sigma \lpar \theta_{1, i})} \; +] \\
& [\sum\limits_{i \in I_{2}, \act(\theta_{2, i}) \notin L} \ell_{\rm brs}(\sigma \rpar \theta_{2,
i})_{\upd(\ddot{E}, \sigma \rpar \theta_{2, i})} \, . \, e\ell_{\rm
brs}^{\sigma}(\toinitial(\widetilde{P}_{1}), \widetilde{P}''_{2, i}, L)_{\upd(\ddot{E}, \sigma \rpar
\theta_{2, i})} \; +] \\
& [\sum\limits_{i \in I_{1}, \act(\theta_{1, i}) \in L} \; \sum\limits_{j \in I_{2}, \act(\theta_{2, j}) =
\act(\theta_{1, i})} \hspace{-1.0cm} \ell_{\rm brs}(\sigma \langle \theta_{1, i}, \theta_{2, j}
\rangle)_{\upd(\ddot{E}, \sigma \langle \theta_{1, i}, \theta_{2, j} \rangle)} \, . \, e\ell_{\rm
brs}^{\sigma}(\widetilde{P}''_{1, i}, \widetilde{P}''_{2, j}, L))_{\upd(\ddot{E}, \sigma \langle \theta_{1,
i}, \theta_{2, j} \rangle)}] \\
\end{array}}
where each of the last four subprocesses on the right is an initial process needed by the forward-reverse
semantics, with the first one being present only if $\act(\theta_{2}) \notin L$ and the presence of the
subsequent three respectively depending on the presence of $\widetilde{P}''_{1}$, $\widetilde{P}''_{2}$, or
both.

\item The subcase $\act(\theta_{2}) \notin L \land (\act(\theta_{1}) \in L \lor \sigma \rpar \theta_{2}
\le_{\dag} \sigma \lpar \theta_{1})$ is like the previous one.

\item If $\act(\theta_{1}) = \act(\theta_{2}) \in L$, let $e\ell_{\rm brs}^{\sigma}(\widetilde{P}_{1},
\widetilde{P}_{2}, L)_{E}$
\cws{0}{\hspace*{-1.5cm}\begin{array}{rl}
= \!\!\! & \ell_{\rm brs}(\sigma \langle \theta_{1}, \theta_{2}
\rangle)^{\color{magenta}{\dag}}_{\upd(\ddot{E}, \sigma \langle \theta_{1}, \theta_{2} \rangle)} \, . \,
e\ell_{\rm brs}^{\sigma}(\widetilde{P}'_{1}, \widetilde{P}'_{2}, L))_{E} \; + \\
& [\sum\limits_{i \in I_{1}, \act(\theta_{1, i}) \notin L} \ell_{\rm brs}(\sigma \lpar \theta_{1,
i})_{\upd(\ddot{E}, \sigma \lpar \theta_{1, i})} \, . \, e\ell_{\rm brs}^{\sigma}(\widetilde{P}''_{1, i},
\toinitial(\widetilde{P}_{2}), L)_{\upd(\ddot{E}, \sigma \lpar \theta_{1, i})} \; +] \\
& [\sum\limits_{i \in I_{2}, \act(\theta_{2, i}) \notin L} \ell_{\rm brs}(\sigma \rpar \theta_{2,
i})_{\upd(\ddot{E}, \sigma \rpar \theta_{2, i})} \, . \, e\ell_{\rm
brs}^{\sigma}(\toinitial(\widetilde{P}_{1}), \widetilde{P}''_{2, i}, L)_{\upd(\ddot{E}, \sigma \rpar
\theta_{2, i})} \; +] \\
& [\sum\limits_{i \in I_{1}, \act(\theta_{1, i}) \in L} \; \sum\limits_{j \in I_{2}, \act(\theta_{2, j}) =
\act(\theta_{1, i})} \hspace{-1.0cm} \ell_{\rm brs}(\sigma \langle \theta_{1, i}, \theta_{2, j}
\rangle)_{\upd(\ddot{E}, \sigma \langle \theta_{1, i}, \theta_{2, j} \rangle)} \, . \, e\ell_{\rm
brs}^{\sigma}(\widetilde{P}''_{1, i}, \widetilde{P}''_{2, j}, L))_{\upd(\ddot{E}, \sigma \langle \theta_{1,
i}, \theta_{2, j} \rangle)}] \\
\end{array}}
where each of the last three summation-shaped subprocesses on the right is an initial process needed by the
forward-reverse semantics, with their presence respectively depending on the presence of
$\widetilde{P}''_{1}$, $\widetilde{P}''_{2}$, or both.
\fullbox

			\end{itemize}

		\end{itemize}

	\end{definition}

	\begin{example}\label{ex:brs_enc_non_seq_proc}

Let $P$ be $P_{1} \pco{\emptyset} P_{2}$, where $P_{1}$ and $P_{2}$ are the initial sequential processes $a
\, . \, \nil$ and $b \, . \, \nil$ so that $\widetilde{P}_{1} = \ell_{\rm brs}(a)_{a^{\dag} . \, \nil} \, .
\, \widetilde{\nil}$ and $\widetilde{P}_{2} = \ell_{\rm brs}(b)_{b^{\dag} . \, \nil} \, . \,
\widetilde{\nil}$. Then:
\cws{0}{\begin{array}{rcl}
\widetilde{P} \: = \: \ell_{\rm brs}^{\varepsilon}(P)_{P} & \!\!\! = \!\!\! & e\ell_{\rm
brs}^{\varepsilon}(\widetilde{P}_{1}, \widetilde{P}_{2}, \emptyset)_{P} \\
& \!\!\! = \!\!\! & \ell_{\rm brs}(\lpar a)_{a^{\dag} . \, \nil \pco{\emptyset} b \, . \, \nil} \, . \,
e\ell_{\rm brs}^{\varepsilon}(\widetilde{\nil}, \widetilde{P}_{2}, \emptyset)_{a^{\dag} . \, \nil
\pco{\emptyset} b \, . \, \nil} \; + \\
& & \ell_{\rm brs}(\rpar b)_{a \, . \, \nil \pco{\emptyset} b^{\dag} . \, \nil} \, . \, e\ell_{\rm
brs}^{\varepsilon}(\widetilde{P}_{1}, \widetilde{\nil}, \emptyset)_{a \, . \, \nil \pco{\emptyset} b^{\dag}
. \, \nil} \\
& \!\!\! = \!\!\! & \lap a, \{ a \} \rap \, . \, \ell_{\rm brs}(\rpar b)_{a^{\dag} . \, \nil \pco{\emptyset}
b^{\dag} . \, \nil} \, . \, e\ell_{\rm brs}^{\varepsilon}(\widetilde{\nil}, \widetilde{\nil},
\emptyset)_{a^{\dag} . \, \nil \pco{\emptyset} b^{\dag} . \, \nil} \; + \\
& & \lap b, \{ b \} \rap \, . \, \ell_{\rm brs}(\lpar a)_{a^{\dag} . \, \nil \pco{\emptyset} b^{\dag} . \,
\nil} \, . \, e\ell_{\rm brs}^{\varepsilon}(\widetilde{\nil}, \widetilde{\nil}, \emptyset)_{a^{\dag} . \,
\nil \pco{\emptyset} b^{\dag} . \, \nil} \\
& \!\!\! = \!\!\! & \lap a, \{ a \} \rap \, . \, \lap b, {\color{red}{\{ a, b \}}} \rap \, . \, \nil + \lap
b, \{ b \} \rap \, . \, \lap a, {\color{red}{\{ a, b \}}} \rap \, . \, \nil \\
\end{array}}
which is different from the encoding of $a \, . \, b \, . \, \nil + b \, . \, a \, . \, \nil$ shown in
Example~\ref{ex:brs_enc_seq_proc}, unless $a = b$ as in that case the backward ready set $\{ a, b \}$
collapses to $\{ a \}$. \\
If instead $P_{1}$ is the non-initial sequential process $a^{\dag} . \, \nil$ and $P_{2}$ is the initial
sequential process $b \, . \, \nil$, so that $\widetilde{P}_{1} = \ell_{\rm brs}(a)^{\dag}_{a^{\dag} . \,
\nil} \, . \, \widetilde{\nil}$ and $\widetilde{P}_{2} = \ell_{\rm brs}(b)_{b^{\dag} . \, \nil} \, . \,
\widetilde{\nil}$, then:
\cws{0}{\begin{array}{rcl}
\widetilde{P} \: = \: \ell_{\rm brs}^{\varepsilon}(P)_{P} & \!\!\! = \!\!\! & e\ell_{\rm
brs}^{\varepsilon}(\widetilde{P}_{1}, \widetilde{P}_{2}, \emptyset)_{P} \\
& \!\!\! = \!\!\! & \ell_{\rm brs}(\lpar a)^{\dag}_{a^{\dag} . \, \nil \pco{\emptyset} b \, . \, \nil} \, .
\, e\ell_{\rm brs}^{\varepsilon}(\widetilde{\nil}, \widetilde{P}_{2}, \emptyset)_{P} \; + \\
& & \ell_{\rm brs}(\rpar b)_{a \, . \, \nil \pco{\emptyset} b^{\dag} . \, \nil} \, . \, e\ell_{\rm
brs}^{\varepsilon}(\ell_{\rm brs}(a)_{a^{\dag} .  \, \nil} \, . \, \widetilde{\nil}, \widetilde{\nil},
\emptyset)_{a \, . \, \nil \pco{\emptyset} b^{\dag} . \, \nil} \\
& \!\!\! = \!\!\! & \lap a^{\dag}, \{ a \} \rap \, . \, \ell_{\rm brs}(\rpar b)_{a^{\dag} . \, \nil
\pco{\emptyset} b^{\dag} . \, \nil} \, . \, e\ell_{\rm brs}^{\varepsilon}(\widetilde{\nil},
\widetilde{\nil}, \emptyset)_{a^{\dag} . \, \nil \pco{\emptyset} b^{\dag} . \, \nil} \; + \\
& & \lap b, \{ b \} \rap \, . \, \ell_{\rm brs}(\lpar a)_{a^{\dag} . \, \nil \pco{\emptyset} b^{\dag} . \,
\nil} \, . \, e\ell_{\rm brs}^{\varepsilon}(\widetilde{\nil}, \widetilde{\nil}, \emptyset)_{a^{\dag} .  \,
\nil \pco{\emptyset} b^{\dag} . \, \nil} \\
& \!\!\! = \!\!\! & \lap a^{\dag}, \{ a \} \rap \, .  \, \lap b, \{ a, b \} \rap \, . \, \nil + \lap b, \{ b
\} \rap \, . \, \lap a, \{ b, a \} \rap \, . \, \nil \\
\end{array}}
If finally $P_{1}$ is the non-initial sequential process $a^{\dag} . \, \nil$ and $P_{2}$ is the non-initial
sequential process $b^{\dag} . \, \nil$, \linebreak so that $\widetilde{P}_{1} = \ell_{\rm
brs}(a)^{\dag}_{a^{\dag} . \, \nil} \, . \, \widetilde{\nil}$ and $\widetilde{P}_{2} = \ell_{\rm
brs}(b)^{\dag}_{b^{\dag} . \, \nil} \, . \, \widetilde{\nil}$, then for $\lpar a \le_{\dag} \! \rpar b$:
\cws{10}{\begin{array}{rcl}
\widetilde{P} \: = \: \ell_{\rm brs}^{\varepsilon}(P)_{P} & \!\!\! = \!\!\! & e\ell_{\rm
brs}^{\varepsilon}(\widetilde{P}_{1}, \widetilde{P}_{2}, \emptyset)_{P} \\
& \!\!\! = \!\!\! & \ell_{\rm brs}(\lpar a)^{\dag}_{a^{\dag} . \, \nil \pco{\emptyset} b \, . \, \nil} \, .
\, e\ell_{\rm brs}^{\varepsilon}(\widetilde{\nil}, \widetilde{P}_{2}, \emptyset)_{P} \; + \\
& & \ell_{\rm brs}(\rpar b)_{a \, . \, \nil \pco{\emptyset} b^{\dag} . \, \nil} \, . \, e\ell_{\rm
brs}^{\varepsilon}(\ell_{\rm brs}(a)_{a^{\dag} .  \, \nil} \, . \, \widetilde{\nil}, \widetilde{\nil},
\emptyset)_{a \, . \, \nil \pco{\emptyset} b^{\dag} . \, \nil} \\
& \!\!\! = \!\!\! & \lap a^{\dag}, \{ a \} \rap \, . \, \ell_{\rm brs}(\rpar b)^{\dag}_{a^{\dag} . \, \nil
\pco{\emptyset} b^{\dag} . \, \nil} \, . \, e\ell_{\rm brs}^{\varepsilon}(\widetilde{\nil},
\widetilde{\nil}, \emptyset)_{a^{\dag} . \, \nil \pco{\emptyset} b^{\dag} . \, \nil} \; + \\
& & \lap b, \{ b \} \rap \, . \, \ell_{\rm brs}(\lpar a)_{a^{\dag} . \, \nil \pco{\emptyset} b^{\dag} . \,
\nil} \, . \, e\ell_{\rm brs}^{\varepsilon}(\widetilde{\nil}, \widetilde{\nil}, \emptyset)_{a^{\dag} .  \,
\nil \pco{\emptyset} b^{\dag} . \, \nil} \\
& \!\!\! = \!\!\! & \lap a^{\dag}, \{ a \} \rap \, . \, \lap b^{\dag}, \{ a, b \} \rap \, . \, \nil + \lap
b, \{ b \} \rap \, . \, \lap a, \{ b, a \} \rap \, . \, \nil \\
\end{array}}
\fullbox

	\end{example}

We now investigate the correctness of the $\ell_{\rm brs}$-encoding. After some compositionality properties,
\linebreak we show that the encoding preserves initiality and -- to a large extent -- backward ready sets.

	\begin{lemma}\label{lem:brs_enc_comp}

Let $a \in A$ and $P, P_{1}, P_{2} \in \procs$ be such that $a \, . \, P, P_{1} + P_{2} \in \procs$. Then:

		\begin{enumerate}

\item $\widetilde{a \, . \, P} = \lap a, \{ a \} \rap \, . \, \widetilde{P}$.

\item $\widetilde{a^{\dag} . \, P} = \lap a^{\dag}, \brs(a^{\dag} . \, P) \rap \, . \, \widetilde{P}$, with
$\brs(a^{\dag} . \, P) = \{ a \}$ if $P$ is initial.

\item $\widetilde{P_{1} + P_{2}} = \widetilde{P}_{1} + \widetilde{P}_{2}$.
\fullbox

		\end{enumerate}

	\end{lemma}

	\begin{proposition}\label{prop:brs_enc_pres_init_brs}

Let $P \in \procs$. Then:

		\begin{enumerate}

\item $\initial(\widetilde{P})$ iff $\initial(P)$.

\item $\brs(\widetilde{P}) = \brs(P)$ if $P$ has no subprocesses of the form $P_{1} \pco{L} P_{2}$ such that
$P_{1}$ and $P_{2}$ are non-initial and the last executed action $b_{1}^{\dag}$ in $\widetilde{P}_{1}$ is
different from the last executed action $b_{2}^{\dag}$ in $\widetilde{P}_{2}$ with $b_{1}, b_{2} \notin L$.
\fullbox

		\end{enumerate}

	\end{proposition}

As an example, for $P$ given by $a^{\dag} . \, \nil \pco{\emptyset} b^{\dag} . \, \nil$ we have that
$\widetilde{P} = \lap a^{\dag}, \{ a \} \rap \, . \, \lap b^{\dag}, \{ a, b \} \rap \, . \, \nil +
\linebreak \lap b, \{ b \} \rap \, . \, \lap a, \{ a, b \} \rap \, . \, \nil$ when the last executed actions
satisfy $\lpar a \le_{\dag} \! \rpar b$ (see end of Example~\ref{ex:brs_enc_non_seq_proc}), hence $\brs(P) =
\{ a, b \}$ but $\brs(\widetilde{P}) = \{ b \}$ for $a \neq b$. However, in $\widetilde{P}$ the backward
ready set $\{ a, b \}$ occurs next to the last executed action $b^{\dag}$, hence it will label the related
transition in $\! \arrow{}{\rm brs} \!$ (see Table~\ref{tab:brs_semantics}). Indeed, the $\ell_{\rm
brs}$-encoding is correct in the following sense.

	\begin{theorem}\label{thm:brs_enc_pres_trans}

Let $P, P' \in \procs$ and $\theta \in \Theta$. Then $P \arrow{\theta}{} P'$ iff $\widetilde{P}
\xarrow{\ell_{\rm brs}(\theta)_{P'}}{\rm brs} \widetilde{P}'$.
\fullbox

	\end{theorem}

	\begin{corollary}\label{cor:brs_enc_pres_bisim}

Let $P_{1}, P_{2} \in \procs$ and $B \in \{ {\rm RB}, {\rm FRB} \}$. Then $P_{1} \sbis{B} P_{2}$ iff
$\widetilde{P}_{1} \sbis{B:\ell_{\rm brs}} \widetilde{P}_{2}$.
\fullbox

	\end{corollary}

	\begin{table}[t]

\[\begin{array}{|lrcll|}
\hline
& & & & \\[-0.4cm]
(\cala_{{\rm R}, 1}) & \widetilde{(P + Q) + R} & \!\!\! = \!\!\! & \widetilde{P + (Q + R)} & \textrm{where
at least two among $P$, $Q$, $R$ are initial} \\
(\cala_{{\rm R}, 2}) & \widetilde{P + Q} & \!\!\! = \!\!\! & \widetilde{Q + P} & \textrm{where at least one
between $P$ and $Q$ is initial} \\
(\cala_{{\rm R}, 3}) & \widetilde{a \, . \, P} & \!\!\! = \!\!\! & \widetilde{P} & \textrm{where
$\initial(P)$} \\
(\cala_{{\rm R}, 4}) & \widetilde{P + Q} & \!\!\! = \!\!\! & \widetilde{P} & \textrm{if $\initial(Q)$} \\
(\cala_{{\rm R}, 5}) & \widetilde{P_{1} \pco{L} P_{2}} & \!\!\! = \!\!\! & e\ell_{\rm
brs}^{\varepsilon}(\widetilde{P}_{1}, \widetilde{P}_{2}, L)_{P_{1} \pco{L} P_{2}} & \textrm{with $P_{k}$ in
R-nf for $k \in \{ 1, 2 \}$} \\
\hline
& & & & \\[-0.4cm]
(\cala_{{\rm FR}, 1}) & \widetilde{(P + Q) + R} & \!\!\! = \!\!\! & \widetilde{P + (Q + R)} & \textrm{where
at least two among $P$, $Q$, $R$ are initial} \\
(\cala_{{\rm FR}, 2}) & \widetilde{P + Q} & \!\!\! = \!\!\! & \widetilde{Q + P} & \textrm{where at least one
between $P$ and $Q$ is initial} \\
(\cala_{{\rm FR}, 3}) & \widetilde{P + \nil} & \!\!\! = \!\!\! & \widetilde{P} & \\
(\cala_{{\rm FR}, 4}) & \widetilde{P + Q} & \!\!\! = \!\!\! & \widetilde{P} & \textrm{if $\initial(Q) \land
\toinitial(P) = Q$} \\
(\cala_{{\rm FR}, 5}) & \widetilde{P_{1} \pco{L} P_{2}} & \!\!\! = \!\!\! & e\ell_{\rm
brs}^{\varepsilon}(\widetilde{P}_{1}, \widetilde{P}_{2}, L)_{P_{1} \pco{L} P_{2}} & \textrm{with $P_{k}$ in
FR-nf for $k \in \{ 1, 2 \}$} \\
\hline
\end{array}\]

\caption{Axioms characterizing $\sbis{\rm RB}$ and $\sbis{\rm FRB}$ via the $\ell_{\rm brs}$-encoding into
$\procs_{\rm brs}$ processes}
\label{tab:rb_frb_axioms}

	\end{table}

The set $\cala_{\rm R}$ of axioms for $\sbis{\rm RB}$ is shown in the upper part of
Table~\ref{tab:rb_frb_axioms}. All the axioms apart from the last one come from the axiomatization developed
in~\cite{BR23} over sequential processes. Axiom~$\cala_{{\rm R}, 3}$ establishes that the future can be
completely canceled when moving only backward. Likewise, axiom~$\cala_{{\rm R}, 4}$ states that a previously
non-selected alternative can be discarded when moving only backward; note that this axiom subsumes both
$\widetilde{P + \nil} = \widetilde{P}$ and $\widetilde{P + P} = \widetilde{P}$. The new axiom~$\cala_{{\rm
R}, 5}$ concisely expresses via $e\ell_{\rm brs}$ the expansion laws for reverse bisimilarity, where $P_{k}$
is $\nil$ or the $+$-free sequential process $a_{k}^{\dag} \, . \, P'_{k}$ featuring only executed actions
for $k \in \{ 1, 2 \}$.

	\begin{definition}\label{def:r_normal_form}

We say that $P \in \procs$ is in \emph{reverse normal form}, written \emph{R-nf}, iff it is equal to $\nil$
or $a^{\dag} . \, P'$ where $P'$ is in R-nf. This extends to $\widetilde{P} \in \procs_{\rm brs}$ in the
expected way.
\fullbox

	\end{definition}

	\begin{lemma}\label{lem:reduc_rnf}

For all (initial) $P \in \procs$ there exists (an initial) $\widetilde{Q} \in \procs_{\rm brs}$ in R-nf
(which is $\widetilde{\nil}$) such that $\cala_{\rm R} \vdash \widetilde{P} = \widetilde{Q}$.
\fullbox

	\end{lemma}

	\begin{theorem}\label{thm:rb_axioms}

Let $P_{1}, P_{2} \in \procs$. Then $\widetilde{P}_{1} \sbis{{\rm RB}:\ell_{\rm brs}} \widetilde{P}_{2}$ iff
$\cala_{\rm R} \vdash \widetilde{P}_{1} = \widetilde{P}_{2}$.
\fullbox

	\end{theorem}

The set $\cala_{\rm FR}$ of axioms for $\sbis{\rm FRB}$ is shown in the lower part of
Table~\ref{tab:rb_frb_axioms}. All the axioms apart from the last one come from the axiomatization developed
in~\cite{BR23} over sequential processes. Axiom~$\cala_{{\rm FR}, 4}$ is a variant of idempotency appeared
for the first time in~\cite{LP21}, with $P$ and $Q$ coinciding like in axiom~$\cala_{{\rm F}, 4}$ when they
are both initial. The new axiom~$\cala_{{\rm FR}, 5}$ concisely expresses via $e\ell_{\rm brs}$ the
expansion laws for forward-reverse bisimilarity, where $P_{k}$ is the sequential process $[a_{k}^{\dag} \, .
\, P'_{k} \, +] \sum_{i \in I_{k}} a_{k, i} \, . \, P_{k, i}$ for $k \in \{ 1, 2 \}$.

	\begin{definition}\label{def:fr_normal_form}

We say that $P \in \procs$ is in \emph{forward-reverse normal form}, written \emph{FR-nf}, iff it is equal
to $[b^{\dag} . \, P' \, +] \sum_{i \in I} a_{i} \, . \, P_{i}$ where $b^{\dag} . \, P'$ is optional, $P'$
is in FR-nf, $I$ is a finite index set (with the summation being $\nil$ -- or disappearing in the presence
of $b^{\dag} . \, P'$ -- when $I = \emptyset$), and each $P_{i}$ is initial and in FR-nf. \linebreak This
extends to $\widetilde{P} \in \procs_{\rm brs}$ in the expected way.
\fullbox

	\end{definition}

	\begin{lemma}\label{lem:reduc_frnf}

For all (initial) $P \in \procs$ there exists (an initial) $\widetilde{Q} \in \procs_{\rm brs}$ in FR-nf
such that $\cala_{\rm FR} \vdash \widetilde{P} = \widetilde{Q}$.
\fullbox

	\end{lemma}

	\begin{theorem}\label{thm:frb_axioms}

Let $P_{1}, P_{2} \in \procs$. Then $\widetilde{P}_{1} \sbis{{\rm FRB}:\ell_{\rm brs}} \widetilde{P}_{2}$
iff $\cala_{\rm FR} \vdash \widetilde{P}_{1} = \widetilde{P}_{2}$.
\fullbox

	\end{theorem}

%%%%%%%%%%%%%%%%%%%%%%%%%%%%%%%%%%%%%%%%%%%%%%%%%%%%%%%%%%%%%%%%%
%
%
\section{Conclusions}
\label{sec:concl}
%
%
%%%%%%%%%%%%%%%%%%%%%%%%%%%%%%%%%%%%%%%%%%%%%%%%%%%%%%%%%%%%%%%%%

In this paper we have exhibited expansion laws for forward bisimilarity, which is interleaving, and reverse
and forward-reverse bisimilarities, which are truly concurrent. To uniformly develop them, we have resorted
to the proved trees approach of~\cite{DP92}, which has turned out to be effective also in our setting. With
respect to other truly concurrent semantics to which the approach was applied, such as causal and location
bisimilarities, the operational semantics of our reversible calculus does not carry the additional
discriminating information within transition labels. However, we have been able to derive it from those
labels and shown to consist of backward ready sets for both reverse and forward-reverse bisimilarities.
Another technical difficulty that we have faced is the encoding of concurrent processes in which both
subprocesses have executed non-synchronizing actions, because their expansions cannot contain executed
actions on both sides of an alternative composition. For completeness we mention that in~\cite{Aub22} proved
semantics has already been employed in a reversible setting, for a different purpose though.

As for future work, an obvious direction is to exploit the same approach to find out expansion laws for the
weak versions of forward, reverse, and forward-reverse bisimilarities, i.e., their versions capable of
abstracting from $\tau$-actions~\cite{BE23a}.

A more interesting direction is to show that forward-reverse bisimilarity augmented with a clause for
backward ready \emph{multi}sets equality corresponds to hereditary history-preserving
bisimilarity~\cite{Bed91}, thus yielding for the latter an operational characterization, an axiomatization
alternative to~\cite{FL05}, and logical characterizations alternative to~\cite{PU14,BC14}. These two
bisimilarities were shown to coincide in~\cite{Bed91,PU07b,PU12,AC17} in the absence of autoconcurrency. In
fact, if $a = b$ in Figure~\ref{fig:expansion_law}, the two processes turn out to be forward-reverse
bisimilar, with the backward ready sets of the three bottom states collapsing to $\{ a \}$, but not
hereditary history-preserving bisimilar, because \emph{identifying} executed actions is
important~\cite{AC20} (as done also in CCSK via communication keys~\cite{PU07a}). However, if backward ready
multisets are used instead, then the bottom state on the left can be distinguished from the two bottom
states on the right. Thus, \emph{counting} executed actions that label incoming transitions may be enough.

\medskip
\noindent
\textbf{Acknowledgments.}
We would like to thank Irek Ulidowski, Ilaria Castellani, and Pierpaolo Degano for the valuable discussions.
This research has been supported by the PRIN 2020 project \emph{NiRvAna -- Noninterference and Reversibility
Analysis in Private Blockchains}, the PRIN 2022 project \emph{DeKLA -- Developing Kleene Logics and Their
Applications}, and the INdAM-GNCS 2024 project \emph{MARVEL -- Modelli Composizionali per l'Analisi di
Sistemi Reversibili Distribuiti}.

%%%%%%%%%%%%%%%%%%%%%%%%%%%%%%%%%%%%%%%%%%%%%%%%%%%%%%%%%%%%%%%%%
%                                                               %
%                                                               %
% References                                                    %
%                                                               %
%                                                               %
%%%%%%%%%%%%%%%%%%%%%%%%%%%%%%%%%%%%%%%%%%%%%%%%%%%%%%%%%%%%%%%%%
\bibliographystyle{eptcs}
\bibliography{biblio}

\end{document}